\documentclass[acmsmall]{acmart}

\usepackage[normalem]{ulem}
\usepackage[utf8]{inputenc}
\usepackage{amsmath}  
\usepackage{graphicx}
\usepackage{comment}
\usepackage{multirow}
\usepackage{enumitem}
\usepackage{array}
\usepackage{balance}
\usepackage{mathtools}
\usepackage{xspace}
\usepackage{url}
\usepackage{eucal}
\usepackage{color}
\usepackage{booktabs}
\usepackage{float}
\restylefloat{table}
\usepackage{xcolor}
\usepackage{subfloat}
\usepackage[captionskip=1pt,farskip=1pt,position=below]{subfig}

\usepackage{moreverb}
\usepackage{fontenc}
\usepackage{fancybox}
\usepackage{colortbl}
\usepackage{natbib}
\usepackage{multicol}
\usepackage{listings}
\usepackage{boxedminipage}
\usepackage[ruled,vlined,lined,commentsnumbered]{algorithm2e}
\usepackage{csvsimple}
\usepackage{longtable}
\usepackage[skip=1pt,labelfont=bf]{caption}
\usepackage{lscape}
\usepackage{rotating}
\usepackage{tikz}
\usepackage{enumitem}
\usepackage{setspace}
\usepackage{hanging}
\usepackage{lipsum} 
\usepackage[most]{tcolorbox}

\usepackage{indentfirst}
\usepackage{threeparttable}

\newenvironment{changemargin}[2]{%
\begin{list}{}{%
\setlength{\topsep}{0pt}%
\setlength{\leftmargin}{#1}%
\setlength{\rightmargin}{#2}%
\setlength{\listparindent}{\parindent}%
\setlength{\itemindent}{\parindent}%
\setlength{\parsep}{\parskip}%
}%
\item[]}{\end{list}}

\newcommand{\benchmark}{{\sc ReName4J}\xspace}

\newcommand{\find}[1]{
\begin{tcolorbox}[tile,size=fbox,boxsep=2mm,boxrule=0pt,top=0pt,bottom=0pt,
borderline west={1mm}{0pt}{blue!50!white},colback=blue!5!white]
\em #1
\end{tcolorbox}
}
\setlist{noitemsep} 

\newcommand{\codeblock}[1]{
\vspace{-7pt}
\begin{tcolorbox}[tile,size=fbox,boxsep=1mm,boxrule=0pt,top=0pt,bottom=0pt,
borderline west={2mm}{0pt}{black!5!white},colback=black!5!white, width=\linewidth] 
\ttfamily #1
\end{tcolorbox}
\vspace{-7pt}
}

\acmBooktitle{ACM TOSEM}

\begin{document}
\title{How are We Detecting Inconsistent Method Names? 
\\ An Empirical Study from Code Review Perspective}


\author{Kisub Kim}
\email{kisubkim@smu.edu.sg}
\affiliation{%
  \institution{Singapore Management University}
  \city{Singapore}
  \country{Singapore}
}

\author{Xin Zhou}
\email{xinzhou.2020@phdcs.smu.edu.sg}
\affiliation{%
  \institution{Singapore Management University}
  \city{Singapore}
  \country{Singapore}
}

\author{Dongsun Kim}
\email{darkrsw@gmail.com}
\affiliation{%
  \institution{Kyungpook National University}
  \city{Daegu}
  \country{Korea}
}

\author{Julia Lawall}
\email{julia.lawall@inria.fr}
\affiliation{%
  \institution{Inria}
  \city{Paris}
  \country{France}
}

\author{Kui Liu}
\email{brucekuiliu@gmail.com}
\affiliation{%
  \institution{Huawei Software Engineering Application Technology Lab}
  \city{Hangzhou}
  \country{China}
}

\author{Tegawend\'e F. Bissyand\'e}
\email{tegawende.bissyande@uni.lu}
\affiliation{%
  \institution{University of Luxembourg}
  \city{Luxembourg}
  \country{Luxembourg}
}

\author{Jacques Klein}
\email{jacques.klein@uni.lu}
\affiliation{%
  \institution{University of Luxembourg}
  \city{Luxembourg}
  \country{Luxembourg}
}

\author{Jaekwon Lee}
\email{jaekwon.lee@uottawa.ca}
\affiliation{%
  \institution{University of Ottawa}
  \city{Ottawa}
  \country{Canada}
}

\author{David Lo}
\email{davidlo@smu.edu.sg}
\affiliation{%
  \institution{Singapore Management University}
  \city{Singapore}
  \country{Singapore}
}


  
\renewcommand{\shortauthors}{Kim et al.}


\begin{abstract}
\label{sec.abstract}
Proper naming of methods can make program code easier to understand, and thus enhance software maintainability. Yet, developers may use inconsistent names due to poor communication or a lack of familiarity with conventions within the software development lifecycle. To address this issue, much research effort has been invested into building automatic tools that can check for method name inconsistency and recommend consistent names. However, existing datasets generally do not provide precise details about why a method name was deemed improper and required to be changed. Such information can give useful hints on how to improve the recommendation of adequate method names. Accordingly, we construct a sample method-naming benchmark, ReName4J, by matching name changes with code reviews. We then present an empirical study on how state-of-the-art techniques perform in detecting or recommending consistent and inconsistent method names based on ReName4J. The main purpose of the study is to reveal a different perspective based on reviewed names rather than proposing a complete benchmark. We find that the existing techniques underperform on our review-driven benchmark, both in inconsistent checking and the recommendation. We further identify potential biases in the evaluation of existing techniques, which future research should consider thoroughly. 

\end{abstract}


\maketitle
``\textit{There are only two hard things in Computer Science: cache invalidation and naming things.}'' —- Philip Lewis Karlton~\cite{karlton}

\section{Introduction}
\label{sec.introduction}

A method is the smallest unit of program behavior in a software project~\cite{host2009debugging,cognac_2021}.  
Accordingly, choosing method names that are consistent with the conventions of the software project and the behavior of the method body is crucial for software maintenance~\cite{arnaoudova2014repent,fakhoury2018effect}, comprehension~\cite{takang1996effects, lawrie2006s,schankin2018descriptive,hofmeister2017shorter}, and reuse~\cite{butler2010exploring,butler2011improving,arnaoudova2016linguistic,liblit2006cognitive}.  
Nevertheless, software developers are not always successful in choosing appropriate method names. 
Figure~\ref{fig:motivating_example} shows a few examples from open-source projects.  
In Figure~\ref{fig:eq1}, the method name {\tt register} suggests that the method stores some data in an internal container while the method actually creates a new object and returns it.  
Thus, the corresponding code review suggests {\tt generateToken} instead.
In Figure~\ref{fig:eq3}, the method name {\tt transitToRestoreActive} suggests that the method changes a program state selectively, while actually, the method is executed all the time.  
The corresponding review recommends {\tt enforceRestoreActive}.

\begin{figure}[!h]
\begin{center}
    \subfloat[A method from {\tt Jenkins} project and its corresponding review comment.\protect\footnotemark]%
	{\parbox{\linewidth}{%
\begin{lstlisting}[linewidth={\linewidth},frame=tb,basicstyle=\fontsize{6.4}{6.4}\selectfont\ttfamily]
public Token register(...) {
    ...
    try {
        return new Token(dbsUrl, authenticationName, Instant.now());
    } catch (Exception ex) {
        LOGGER.log(...);
    }
    return null;
}
\end{lstlisting}\vspace{-3mm}
    ~\\
    \noindent\fcolorbox{gray}{gray!30}{%
    \begin{minipage}{0.97\linewidth}{\footnotesize
    {\bf register} should be renamed to be consistent with the current approach (was more meaningful when it was stateful). Proposal: {\bf generateToken}.}
    \end{minipage}}%
}\label{fig:eq1}}%
\newline\vspace{3mm}
    \subfloat[A method from {\tt Kafka} project and its corresponding review comment.\protect\footnotemark]%
	{\parbox{\linewidth}{%
\begin{lstlisting}[linewidth={\linewidth},frame=tb,basicstyle=\fontsize{6.4}{6.4}\selectfont\ttfamily]
public void transitToRestoreActive() {
    if (state != ChangelogReaderState.ACTIVE_RESTORING) {
      ...
      pauseChangelogsFromRestoreConsumer(standbyRestoringChangelogs());    
    }
    state = ChangelogReaderState.ACTIVE_RESTORING;
}
\end{lstlisting}\vspace{-3mm}
        ~\\
        \noindent\fcolorbox{gray}{gray!30}{%
        \begin{minipage}{0.97\linewidth}{\footnotesize
        Renamed this method to make it clear we aren't necessarily "transitioning", we actually call it all the time now any time we want to "be in restoreActive".}
        \end{minipage}}%
}\label{fig:eq3}}%
    \caption{Motivating examples taken from {\tt Apache} projects.}
    \label{fig:motivating_example}
\end{center}
\end{figure}

\footnotetext[1]{\url{https://github.com/jenkinsci/jenkins/pull/4239/files/41919ee7829223062d5d5ca4d592fd8056e55017\#r330954855}}
\footnotetext[2]{\url{https://github.com/apache/kafka/pull/8319/files/c7ac051e495a98b6c72a340c24f4b9bb1f25dcdd\#r395348873}}

To alleviate the problem of inconsistent method names, many researchers have proposed techniques~\cite{liu2019learning,nguyen2020suggesting,deepname,allamanis2015suggesting,9712079} to automate two tasks: (1) \underline{\textbf{m}}ethod name \underline{\textbf{c}}onsistency \underline{\textbf{c}}hecking (MCC) and (2) \underline{\textbf{m}}ethod \underline{\textbf{n}}ame \underline{\textbf{r}}ecommendation (MNR).  
A challenge, however, is how to evaluate such approaches in a meaningful way.
For example, the dataset of Liu et al.~\cite{liu2019learning} comes from revisions that only change method names. 
While creating a revision that only changes method names does suggest that the method names were unsuitable, such a revision does not provide clear evidence of whether the change reflects the conventions of a community or the opinion of a single developer. 
Datasets furthermore have been balanced, to meet the needs of learning methods and report results on the recognizability of specific classes.  
But in practice, method names tend to stabilize over time, and thus most methods in a well-maintained software project already have consistent names, so a technique scanning for naming problems has to be able to cope with imbalanced inputs.  
Finally, techniques often rely on similarity thresholds that have been optimized for specific datasets.  
These issues indicate that there may be a gap between the research results and reality.

Code review data can provide insight to address the above method naming challenges.
Alsuhaibani et al.~\cite{alsuhaibani_2021_2021} found that method names are discussed during code review and that developers change method names in response to comments from reviewers.  
Such discussions raise an opportunity to sample method-renaming pairs, in which method name changes are supported by discussions among maintainers of the affected code base, and to use this database to assess MCC and MNR techniques. 
For our benchmark, \benchmark, we first collect the reviews from pull requests in
GitHub~\cite{github}.  
We leverage naming-related terms (e.g., ``naming'',
``name'', ``inconsistent'', and ``descriptive'') to identify relevant
reviews and collect the corresponding patches, including the actual code
files as well as the metadata (e.g., file hashes, paths, pull request
status, and commit information).  To extract the method names, we parse all
the collected patches and files, then we map the extracted names to the
reviews.  
We include 400 sample pairs of method names (i.e., buggy and
fixed), method bodies, and the corresponding reviews.

Based on \benchmark, this paper investigates existing MCC and MNR techniques and asks the following research questions:
\begin{itemize}
    \item RQ1: Can the existing MCC/MNR tools retrieve the correct names for the methods that are considered in code review scenarios?
    \item RQ2: Do the existing MCC/MNR tools achieve an acceptable detection rate for consistent method names as well as for those of inconsistent method names? 
    \item RQ3: What happens when we test the MCC/MNR tools on an imbalanced test dataset reflecting the properties of real world software projects instead of a balanced test dataset that is artificially selected?
\end{itemize}

Answering these questions helps fill the gaps between the research and practice by providing three insights.
First, the effectiveness of existing techniques for measuring MCC/MNR relies on the validity of the potential ground truth.
Second, it is important to consider how current tools fare in both identifying inconsistencies and verifying consistency.
Third, the impact of the balance or imbalance of test sets on performance should also be considered.
We believe that these research questions could capture the attention of researchers who are interested in software refactoring specifically, method name recommendation and consistency checking. 
Moreover, the researchers and practitioners can better comprehend various perspectives to evaluate the relevant tools.

To address these questions, we present an empirical study of how state-of-the-art techniques detect/recommend the names for test oracles from our benchmark.
The study involves three recent techniques, Spot~\cite{liu2019learning}, Cognac~\cite{cognac_2021}, and GTNM~\cite{gtnm_2022}, for MCC/MNR tasks.
Our main findings are as follows:

\begin{itemize}
    \item The datasets employed in previous evaluations encompass methods for which consistency has not undergone comprehensive verification through reviews, thereby lacking the assurance that the method names are validated.
    In contrast, our benchmark includes method names checked and accepted by the project code reviewers.
    By providing the verified method names, the benchmark could be more practical and valuable for more precise evaluation.
    \item The existing test dataset~\cite{liu2019learning} contains
    many methods that have simple names consisting of a verb such as  ``get'', ``take'', and ``listen''.
    These intuitive method names are easy to recommend as method bodies tend to include many such tokens. 
    However, the method names in code reviews explicitly show that reviewers prefer more descriptive and comprehensive names.
    Indeed, the average number of characters for the existing test method names from the existing dataset and those of our benchmark are 13.18 and 24.90 (approx. 88.92\% longer), respectively.
    \item The existing techniques show the performance drops for all the tasks on our human-reviewed method names as compared to results in the original studies.
    \item The evaluation design of existing recommendation-based techniques needs to be carefully re-considered as it may be biased for specific cases.
\end{itemize}

In summary, this paper contributes the following:
\begin{itemize}
    \item The first in-depth investigation of MCC/MNR tools, reflecting a code-review perspective.
    \item 
    A small, but reliable benchmark, consisting of sample pairs of method names (i.e., buggy and fixed), method bodies, and review comments.
    This is also the first manually labeled naming dataset from a code review perspective.
    It may serve as a baseline dataset for future MCC/MNR research as the method names have been confirmed by project reviewers.
    \item 
    Delving into advanced evaluation methods for MCC/MNR techniques due to variations in performance measurement protocols within prior literature. 
    Moreover, examining strategies to construct an improved benchmark that minimizes manual effort.
\end{itemize}

The experimental results show that the review-driven benchmark could provide insights for better naming by exploiting more context available in the reviews from the experts.
The provided reviews in \benchmark also can give intuitions on how to improve the classification performance by utilizing more information in the body of the methods. 
\begin{figure}[h!]
\begin{center}
	\subfloat[A review from Apache/Flink project.\protect\footnotemark]{
		\includegraphics[width=0.9\linewidth]{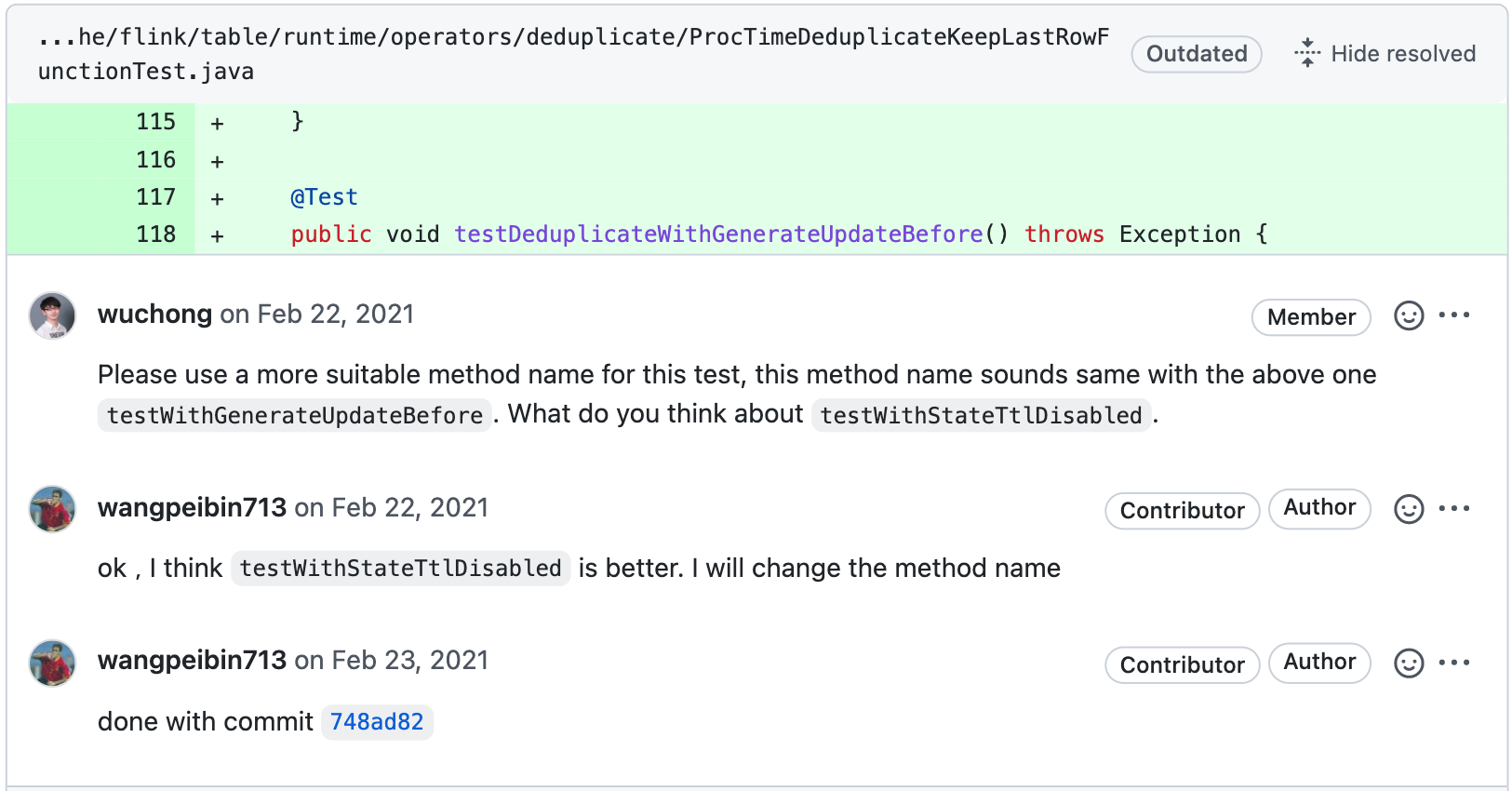}
        \label{subfig:a}
        }
	\\
	\subfloat[A review from Elastic/Elasticsearch project.\protect\footnotemark]{	
            \includegraphics[width=0.9\linewidth]{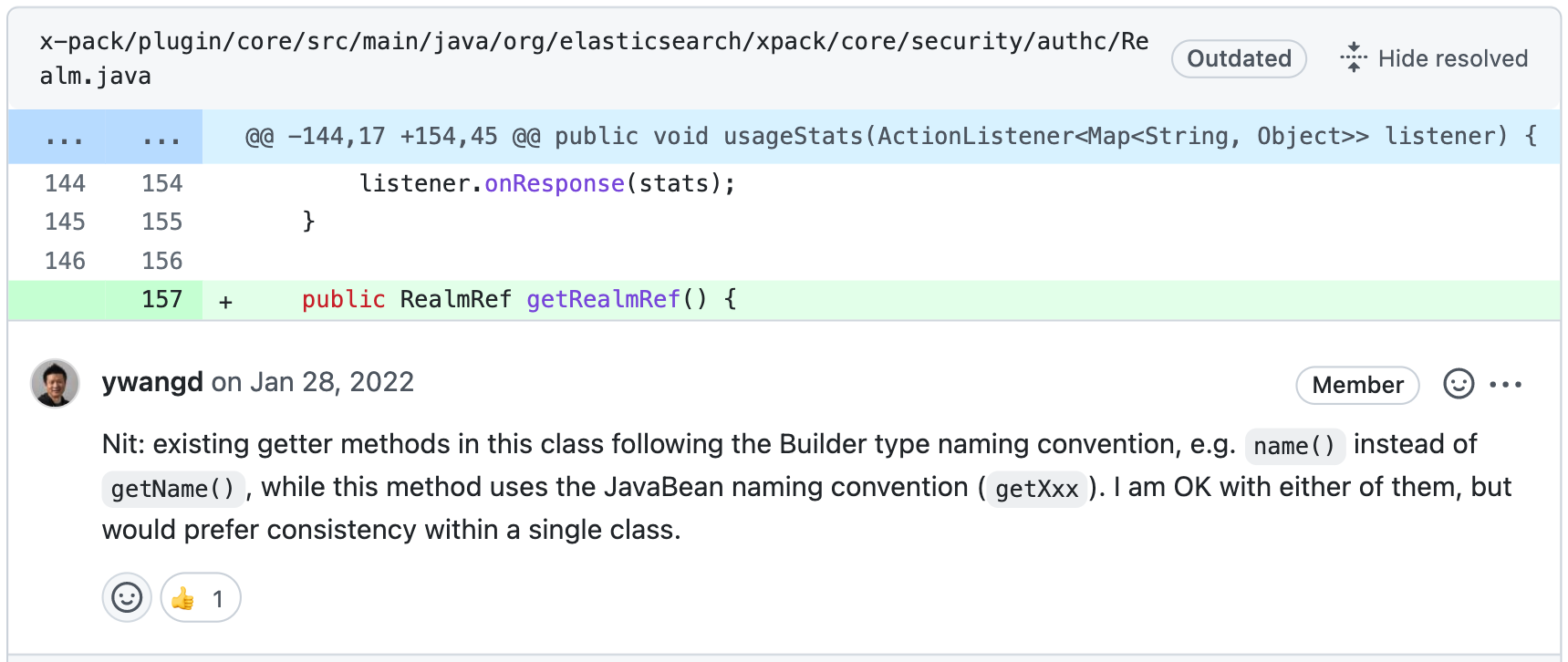}
\label{subfig:b}	
 }
        
        \caption{Real world code reviews on method names.}
\label{fig:review_example}
\end{center}
\end{figure}

\footnotetext[3]{\url{https://github.com/apache/flink/pull/14863}}
\footnotetext[4]{\url{https://github.com/elastic/elasticsearch/pull/82639}}

\section{Motivating Example}
\label{sec:motivation}
The naming of methods is a critical issue for software engineering and development as method names have a high impact on the comprehension of the source code~\cite{takang1996effects,maletic2001supporting}.
Consequently, developers actively discuss naming issues frequently during code reviews in commercial and open-source software projects~\cite{alsuhaibani_2021_2021}.
We first investigate the method names and review pairs in open-source projects to determine whether the explicitly reviewed method names could be a better ground truth for MCC and MNR tasks.


Figure~\ref{fig:review_example} illustrates two examples of naming-related reviews from Apache/Shardingsphere and Elastic/Elasticsearch, respectively.
Figure~\ref{subfig:a} shows a case in which a code reviewer recommends using a more suitable method name for a test while giving the justification that the current name sounds the same as that of another method.
Figure~\ref{subfig:b} illustrates a renaming argument based on the specific naming conventions of the project. 
Specifically, the original name was {\tt getRealmRef()}, which follows a common Java-Bean naming convention, but the reviewer suggested instead following the Builder naming convention, illustrated as {\tt name()}, which implies that the name should be {\tt realmRef()}.

Investigating various examples, we discovered that the code reviewers tend to comment on the method names in many ways such that the developers can directly apply a proposed renaming or be inspired to correct the names.
These changes are explicitly discussed, changed, and accepted.
We consider that these cases are more solid than other renaming cases that are only changed and accepted without any discussion in software development.

\section{Methodology}
\label{sec.methodology}
\begin{figure*}[!ht]
 \begin{center}
  \includegraphics[width=\linewidth]{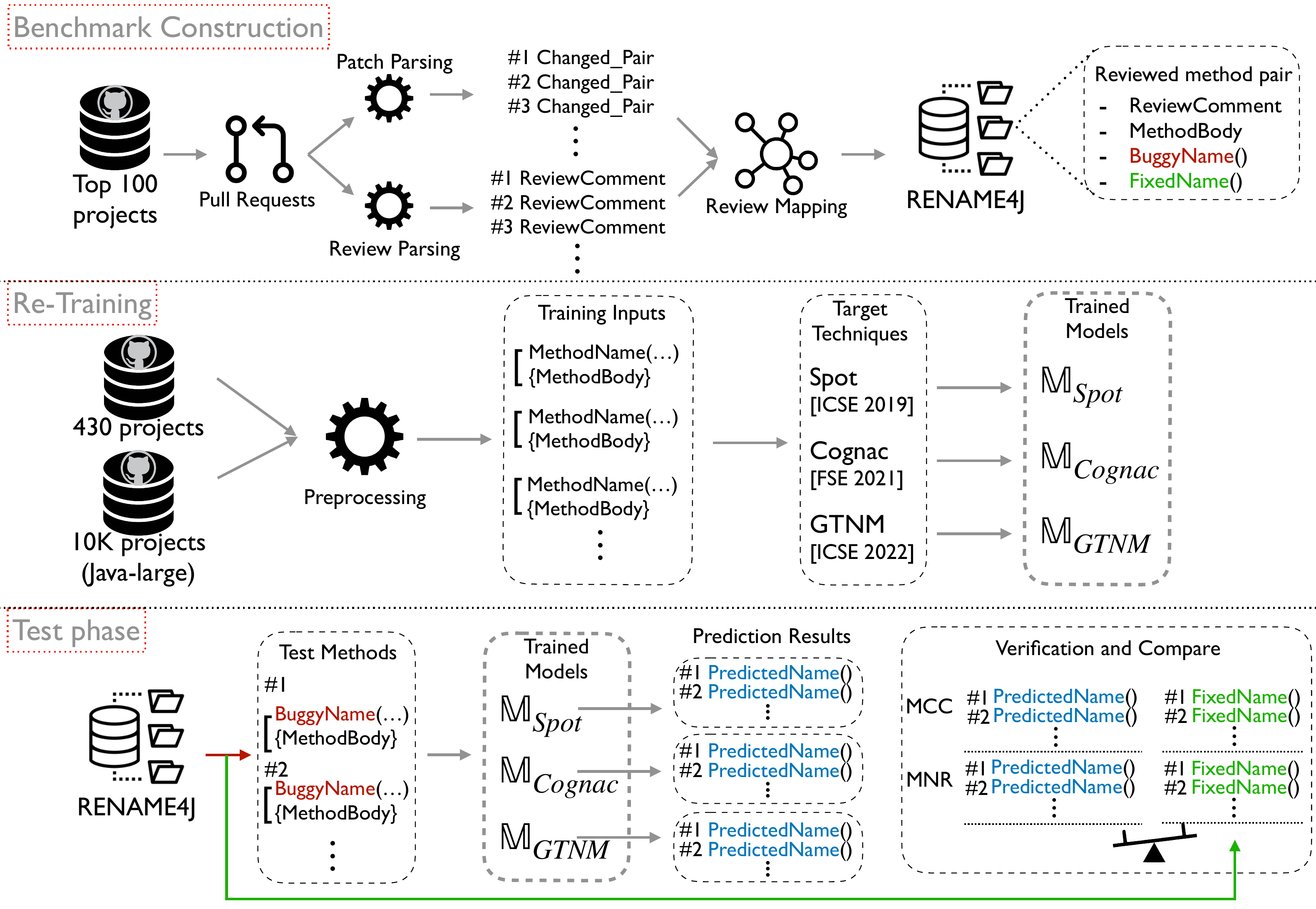}
  \caption{Overall procedure for our study.}
  \label{fig:overview}
 \end{center}
\end{figure*}

This section presents an overview of our empirical study.
As shown in Figure~\ref{fig:overview}, our study comprises three steps:
(1) constructing a novel method naming benchmark (\benchmark), (2) retraining existing method-name recommendation techniques, and (3) testing the performance of the techniques with respect to the benchmark. Our benchmark, \benchmark
collects method renaming instances from code review data such that every instance is backed by discussions of real developers and reviewers.
Based on the benchmark and obtained models, we conduct two tasks: (1) inconsistency classification of each pair of method name and body (MCC), and (2) method name recommendation for a given method body (MNR).

\subsection{Target Techniques}
\label{subsec.subject}
To answer the research questions described in Section~\ref{sec.introduction},
we select three target techniques, Spot~\cite{liu2019learning}, Cognac~\cite{cognac_2021}, and GTNM~\cite{gtnm_2022} as (1) they have been published recently at top software engineering venues
and (2) their implementations are completely reproducible.

\noindent
\textbf{\textit{Spot}}: Spot~\cite{liu2019learning} is designed with the intuition {\it ``Methods implementing similar behavior in their body code are likely to be consistently named with similar names, and vice versa''}.
Notably, Spot adopts unsupervised learning~\cite{hastie2009unsupervised} and lazy learning~\cite{aha2013lazy} to embed method names and bodies into numerical vectors.
Given a method name and body, it compares two sets of names.
According to the embedding vectors, the first set consists of other method names similar to the name of the given method and the second set consists of the names of methods whose bodies are identified as similar to the body of the given method.
When the intersection of these two sets is larger than a threshold, the method name is recognized as consistent; otherwise, Spot classifies it as inconsistent (i.e., the MCC task).
If the name turns out to be inconsistent, Spot recommends potential consistent names for the method based on the comparison results on method bodies (i.e., the MNR task).

\noindent
\textbf{\textit{Cognac}}:
Cognac~\cite{cognac_2021} is a program-structure-independent approach aiming to avoid the out-of-vocabulary problem.
Unlike prior techniques, Cognac not only leverages the \textit{local context} extracted from target methods themselves, including code tokens and the associated contexts, but it also considers the \textit{global context}, which is the contextual information from other methods such as call relations (i.e., caller/callee information) with the target method.
Technically, it follows the \textit{seq2seq} paradigm, using the extracted token sequences to infer method names.

\noindent
\textbf{\textit{GTNM}}:
GTNM~\cite{gtnm_2022} is a transformer-based neural model considering the information of the whole project (global context).
It extracts context from three different levels, given the target method and the project, including (1) local context, (2) project-specific context, and (3) documentation context.
Technically, it employs a transformer-based seq2seq framework~\cite{vaswani2017attention} with two encoders and a single decoder to generate the method name.
Particularly, we include GTNM in our study because its design closely resembles that of Cognac while it misses the evaluation of MCC.
Since Cognac has already showcased that an MNR technique can effectively serve as an MCC task, it is crucial to undertake a comprehensive performance assessment in this context.
This addition could shed light on the potential advantages associated with choosing the right approach for tasks that necessitate a specific tool.

Two other approaches, MNire~\cite{nguyen2020suggesting} and DeepName~\cite{deepname} have been recently proposed and shown to outperform previous approaches.
We had to discard MNire~\cite{nguyen2020suggesting} due to the unavailability of the source code~\cite{cognac_2021}. 
Fortunately, Cognac~\cite{cognac_2021} has been compared against the numbers in the MNire paper with the same dataset, and Cognac outperforms MNire.
We failed to execute DeepName~\cite{deepname} based on its replication package, as the instructions are not complete, and the author who is in charge of the package did not reply to our questions.
Nevertheless, DeepName is similar to Cognac, in that both use caller/callee information, and both were released at almost the same time.
Both Cognac and DeepName were evaluated against MNire, and they show approximately 10\% and 7\% improvement, respectively, in F1-score with the same dataset for the MNR task.

\subsection{\benchmark: Naming Review Benchmark}
\label{subsec.benchmark}
Our study first builds a benchmark based on code review data, which can be used to assess method-name recommendation techniques.
To the best of our knowledge, there is no concrete benchmark for method (re)naming that incorporates code-review data.
Existing techniques are evaluated on a dataset collected from source code revision history without an associated explanation for the naming changes. 
Distinguishing between true renaming tasks and coincidental code changes proves challenging, as developers often disperse a unified modification across multiple commits~\cite{wen2020empirical,wen2022quick}.
In contrast, \benchmark leverages code review data to collect actual method renaming tasks, supported by justifications of why the name should be revised. 
Notably, the dataset could be applied to both traditional and learning-based approaches.

\benchmark has four components for each method renaming task: (1) {\bf ReviewComment}: the review comment explaining why the name should be revised, (2) {\bf MethodBody}: the relevant method body (i.e., source code), (3) {\bf BuggyName}: the method name before refactoring ({\texttt buggy\_method\_name()}), and (4) {\bf FixedName}: the method name after refactoring ({\texttt fixed\_method\_name()}).

To collect the review and method name pair dataset, we leverage GitHub's REST API.\footnote{https://docs.github.com/en/rest}
We first crawl the top 100 Java projects sorted by the stars, considering the project's popularity~\cite{gu2018deep}.
Among the top 100 Java projects, we discard the projects where the discussions are primarily written in languages other than English.
This leaves $66$ projects as our targets.
We then crawled a total of $144,759$ pull requests for these projects.
As we only need to consider the pull requests with naming-related reviews,
we extract only those pull requests that contain naming-related keywords such as ``naming'', ``name'', ``inconsistent'', and ``descriptive''.
Based on the process, we collected $11,301$ pull requests.
A pull request tends to include multiple commits, and a commit can be associated with multiple files.
Our dataset thus covers $73,985$ commits and a total of 593,240 associated files.

We preprocess the collected code review data as follows.
We break down the patch files to discover the methods where only the name changes while excluding those whose names have been changed with their body as changes in the body can naturally affect the name.
This step results in 1,212 changed method names.
Connecting reviews with their corresponding methods is difficult due to the structure of GitHub's database, as it does not allow for direct linking between a review comment and a specific method. Typically, review comments pertain to an entire pull request.
Another obstacle to mapping reviews to methods is that reviewers often comment on external code links that cannot be accessed through GitHub's API but are visible on the web.
Finally, we tackle the issue by establishing a connection between the patch files containing commit modifications and the metadata of pull requests. This metadata encompasses crucial information such as commit hashes, parent hashes, and review comments.
The authors manually investigated all the pairs of reviews and method names and only included those that had a clear connection between the review and the method.
The examples in Section~\ref{sec:motivation} also demonstrate how this connection can be manually labeled by matching the context of the reviews and methods.

The manual checking and filtering criteria are as follows:
\begin{itemize}
    \item We include methods where the pull request was accepted by the code reviewer of the project.

    \item We include methods with names that are directly suggested by the reviewer (e.g., testGetConnectionSession() $\rightarrow$ assertGetConnectionSession() | \textit{Based on the review: Please rename `testGetConnectionSession' to `assertGetConnectionSession'}\footnote{\url{https://github.com/apache/shardingsphere/pull/18618}}).

    \item We include methods with names that are undoubtedly inspired by the review (e.g., completeRestoration() $\rightarrow$ completeRestorationIfPossible() | Based on the review: \textit{nit: can we rename this to `maybeCompleteTaskTypeTransition' or `completeTaskTypeConversionIfNecessary', etc? Right now, it kind of sounds like we just randomly attempt to convert it to a new task type out of nowhere.}\footnote{\url{https://github.com/apache/kafka/pull/8988}}).

    \item We include methods with names that are recommended based on the project's naming convention (e.g., realmRef() $\rightarrow$ getRealmRef() | Based on the review: \textit{Nit: existing getter methods in this class following the Builder type naming convention, e.g., name() instead of getName(), while this method uses the JavaBean naming convention (getXxx). I am OK with either of them but would prefer consistency within a single class}\footnote{\url{https://github.com/elastic/elasticsearch/pull/82639}}).

    \item We include methods with names that are reviewed and commented on for consistency (e.g., register() $\rightarrow$ getToken() | Based on the review: \textit{\textit{`register' should be renamed to be consistent with current approach (was more meaningful when it was stateful)}}\footnote{\url{https://github.com/jenkinsci/jenkins/pull/4239}}).

    \item We ignore simple typos, as they can be noise in the benchmark, as well as example methods.

\end{itemize}

The selected commits satisfy at least one of the defined criteria.
As a result, we identify, as the ground truth, 400 method renaming tasks supported by the corresponding reviews.
In addition, we explicitly eliminate all the class files associated with the corresponding methods from the training data to ensure that there is no data leakage~\cite{ribeiro2016should}.
The absolute number is smaller than the datasets used in the prior studies~\cite{liu2019learning,cognac_2021,gtnm_2022}, but our benchmark collects only obvious renaming tasks confirmed by code reviewers and developers. This allows a more precise evaluation of method renaming techniques.

To understand the characteristic differences between the existing test oracles~\cite{liu2019learning} and those from our benchmark, we further conduct a brief preliminary investigation.
Based on our observation, the average number of characters for the existing test method names and those of our benchmark are 13.18 and 24.90 (approx. 88.92\% longer), respectively, while those of split sub-tokens are 2.52 and 4.53 for each method name.
This indicates a clear difference whereas our benchmark contains method names that are more descriptive, and it suggests that the code reviewers, in practice, prefer to describe as many parts of the methods as possible.


\subsection{Experiment Design}
To conduct a fair comparison, we employ the following procedure~\cite{cognac_2021} for all three target techniques.
Overall, we evaluate the three techniques described in Section~\ref{subsec.subject} with two tasks: Method Name Recommendation (MNR) addressing RQ1 and Method Name Consistency Checking (MCC) addressing RQ2, after retraining them by using the datasets. Then, we feed the method name/body pairs in \benchmark to the (retrained) target techniques and measure the performance with respect to several metrics. In addition, our study evaluates the techniques on more realistic data (i.e., a highly unbalanced dataset including the name/body pairs in \benchmark) to see their effectiveness in a different scenario.


\subsubsection{Retraining}
\label{subsubsec.dataset}

To avoid bias from the dataset, we re-train the recommendation-based target techniques (denoted as $\mathbb{M}$, e.g., $\mathbb{M}_{Cognac}$ and $\mathbb{M}_{GTNM}$) with a single identical dataset.
We leverage the \textit{Java-large} dataset, released by Alon et al.~\cite{alon2018code2seq}, as the training and validation dataset for two, Cognac~\cite{cognac_2021} and GTNM~\cite{gtnm_2022} of our target techniques.
This dataset is the most popular~\cite{alon2018code2seq,alon2019code2vec,nguyen2020suggesting,cognac_2021,deepname,gtnm_2022} and well-maintained for the method name recommendation task.
It consists of 10,222 top-ranked Java projects from GitHub, including 14,458,828 methods and 1,807,913 unique files.
We randomly shuffled and split all the projects into 9,772 training and 450 validation projects.
In addition, to avoid any data leakage, we carefully checked and eliminated all methods included in \benchmark (Section~\ref{subsec.benchmark}) from the training dataset so that there is no intersection between training and testing datasets.
Table~\ref{tab:training_dataset} shows the training and validation data statistics.

\begin{table}[!htp]
 \caption{Statistics of the training and validation dataset.}
 \begin{tabular}{lrrr}
  \hline
           & \textbf{Train} & \textbf{Validation} & \textbf{Total} \\ \hline
  Projects & 9,772          & 450                 & 10,222         \\
  Files    & 1,756,282      & 51,631              & 1,807,913      \\
  Methods  & 13,992,028     & 466,800             & 14,458,828     \\ \hline
 \end{tabular}
 \label{tab:training_dataset}
\end{table}

Exceptionally, we use a different training dataset for the checking-based approach (denoted as $\mathbb{M}_{Spot}$), Spot, as it has an already-known scalability issue that we confirmed with its authors~\cite{liu2019learning}.
As \textit{Java-large} is almost seven times bigger than the dataset originally used for training Spot, the authors claimed that it is impossible to train Spot with the \textit{Java-large} dataset.
Therefore, we utilize their original training dataset to reproduce the model.
The original dataset for Spot includes
2,116,413 methods extracted from 430 Java projects.

\subsubsection{Method Name Recommendation (MNR) Task}
\label{sec:mnr}
The first task to measure the effectiveness of the target techniques is recommending method names for a given method implementation (i.e., method body). In this task, each target technique, $\mathbb{M}$, takes a method body ({\bf MethodBody} in \benchmark) and produces a set of candidate method names that best describe the body, as described in the following equation: 
\begin{equation}
  \mathbb{M}: B^{\ast} \rightarrow N^{\ast}
\end{equation}
\noindent
where $B$ is an alphabet allowed for a method body and $N$ is an alphabet allowed for a method name, respectively. $B^{\ast}$ and $N^{\ast}$ are sets of sequences over $B$ and $N$, respectively. 
After feeding a method body, $b \in B^{\ast}$, available in \benchmark, we produce a recommended name by $\mathbb{M}(b)=n \in N^{\ast}$. We then compare $n$ with our ground truth name, {\bf FixedName}, in the benchmark. The results of this task for each technique are discussed in Section~\ref{sec.rq1} as RQ1.

While Cognac~\cite{cognac_2021} and GTNM~\cite{gtnm_2022} were designed for method name recommendation, Spot~\cite{liu2019learning} was originally built for consistency checking and later added recommendation functionality.
To ensure a fair comparison, we replaced the target method names of our test methods with meaningless tokens when using Spot, which only predicts method names when they are classified as inconsistent.

\subsubsection{Method Name Consistency Checking (MCC) Task}
\label{sec:mcc}

Instead of recommending a name directly, it might be useful if a technique clarifies whether a given method name is inappropriate to describe the method body.
This consistency check can help developers or code reviewers scan a project and figure out the overall naming practice.
Equation~\ref{eq:mcc} represents the MCC task, $\mathbb{D_{\mathbb{M}}}$, for a given method body $b \in B^{\ast}$, method name $n \in B^{\ast}$, and specific name recommendation technique $\mathbb{M}$.
%
\begin{equation}
  \label{eq:mcc}
  \mathbb{D_{\mathbb{M}}}:  B^{\ast} \times N^{\ast} \rightarrow \{C, IC\}
\end{equation}
where $C$ and $IC$ denotes {\it consistent} and {\it inconsistent} name.
Basically, we expect that the verdict should be $C$ when feeding a method body
({\bf MethodBody}) and corresponding {\bf FixedName} in \benchmark. Otherwise,
it should be $IC$ if {\bf BuggyName} is given.
The results of this task are discussed in Section~\ref{sec.rq2} as RQ2.

State-of-the-art techniques~\cite{nguyen2020suggesting,deepname} claimed that it is possible to conduct method name consistency checking based on method name recommendation by employing a specific similarity checking metric,
described in the following subsection with other metrics.
Several prior method name recommendation approaches~\cite{cognac_2021,deepname,gtnm_2022} use this hypothesis to check the consistency of the method names obtained using their recommendation models.


\subsubsection{Reflection of the Real World}
\label{sec:reality}
In the real world, the likelihood of encountering method names that are inconsistent is significantly lower than those that are consistent.
To accurately reflect this situation, we adopt another dataset.
As our target methods in the benchmark are small parts of multiple class files, there exist remaining methods from these files (i.e., the majority of Java class files tend to consist of multiple methods).
The quantity of these remaining methods is substantially greater (13,137) in comparison to our test oracles (400).
These are stable methods that have not been modified, and thus, we consider a name of a method as {\bf StableName} and a body as {\bf StableBody}.
Note that these are not explicitly identified as stable during the code review process, but this is a common assumption as checking every single method that is not reviewed is practically infeasible.
Also, this is the same process with a prior study~\cite{liu2019learning}.
This implies that we can use them to simulate a realistic scenario, and we utilize them for our experimentation in addressing real world reflection (i.e., RQ3), which aims to reflect reality.
To conduct an experiment on such a dataset, we take the same protocol as the MCC task while the verdict should be $C$ when feeding a {\bf StableName}. Otherwise, it should be $IC$ if a {\bf BuggyName} is given.
The results of this task are reported in Section~\ref{sec.rq3} as RQ3.

\subsection{Performance Metrics}
\label{subsec.metrics}

\subsubsection{Metrics for Method Name Recommendation}
\label{sec:mnrmetrics}
As the objective of the MNR task is to recommend the same or similar names to
those that human developers wrote, we compare the ground truth names in the \benchmark with the names generated by the target techniques.
Note that the unit of comparison is tokenized words (i.e., sub-tokens) of the names rather than the whole chunk of the names to capture partial matches.
We employ \textit{Precision}, \textit{Recall}, \textit{F1-score}, and \textit{EMAcc} (Exact Match Accuracy) to evaluate the results of the MNR task, as represented in Equations~\ref{eq:precision}, \ref{eq:recall}, \ref{eq:f1}, and~\ref{eq:emacc}, respectively.

We use the followings notations for the equations: $o \in N^{\ast}$ and $p \in N^{\ast}$
are a ground truth name (i.e., test oracle) in \benchmark and
a name recommended by a targeted technique based on the method body corresponding to $o$, respectively.
$token()$ is a function defined from $N^{\ast}$ to its power set as follows: 
\begin{equation}
  token: N^{\ast} \rightarrow \mathcal{P}(N^{\ast})
\end{equation}
where $token(name)$ produces a token set after tokenizing $name$. This study applies the camel case tokenization as it is the naming convention of Java.

\begin{itemize} [leftmargin=*]
 \item{\bf $\mbox{Precision}_{MNR}$ (positive predictive value)}: is the metric that represents an estimation of how many tokens are correctly predicted within all the predicted tokens. It is expressed as follows:
     \begin{equation}
       \label{eq:precision}
      \mbox{Precision}_{MNR}(o,p) = \frac{\left| token(p)\cap token(o) \right|}{\left| token(p) \right|}
    \end{equation}

\item{\bf $\mbox{Recall}_{MNR}$ (sensitivity)}: is the metric that represents an estimation of how many tokens are correctly predicted within all the oracle tokens. It is expressed as follows:
    \begin{equation}
      \label{eq:recall}
    \mbox{Recall}_{MNR}(o,p) = \frac{\left| token(p)\cap token(o) \right|}{\left| token(o) \right|}
    \end{equation}

\item{\bf $\mbox{F1-score}_{MNR}$}: is the harmonic mean of the precision and recall. It weights the two ratios (precision and recall) in a balanced way:
    \begin{equation}
      \label{eq:f1}
    \mbox{F1-score}_{MNR}(o,p) = \frac{2\times Precision(o,p)\times Recall(o,p)}{Precision(o,p)+Recall(o,p)}
    \end{equation}
\end{itemize}

Furthermore, we take $\mbox{EMAcc}(o,p)$, i.e., Exact Match Accuracy, to assess whether a technique can recommend a name exactly the same as the ground truth name in \benchmark. The function is defined as: 
\begin{equation}
  \label{eq:emacc}
  \mbox{EMAcc}: N^{\ast} \times N^{\ast} \rightarrow \{0, 1\}
\end{equation}
where $\mbox{EMAcc}(o,p)=1$ if $o$ and $p$ are identical.
Otherwise, the value is 0.

\subsubsection{Metrics for Method Name Consistency Checking}
\label{sec:mccmetrics}

%
%


To measure the performance on the MCC task, we use metrics from prior studies~\cite{liu2019learning,cognac_2021,nguyen2020suggesting,deepname}: Precision, Recall, F-score, and Accuracy.
MCC can have four possible outcomes: IC classified as IC (i.e., true positive=TP), IC classified as C (i.e., false negative=FN), C classified as C (i.e., true negative=TN), and C classified as IC (i.e., false positive=FP).
Accordingly, the metrics are defined as follows:
$\mbox{Precision}_{IC} = \frac{\left| TP \right|}{\left| TP \right| + \left| FP \right|}$,
$\mbox{Recall}_{IC} = \frac{\left| TP \right|}{\left| TP \right| + \left| FN \right|}$,
$\mbox{Precision}_{C} = \frac{\left| TN \right|}{\left| TN \right| + \left| FN \right|}$,
$\mbox{Recall}_{C} = \frac{\left| TN \right|}{\left| TN \right| + \left| FP \right|}$,
the {F1-score} is computed as $\frac{2 \times Precision \times Recall}{Precision+Recall}$ for each class.
Moreover, the $Accuracy$ on the whole dataset is defined as $\frac{\left| TP \right| + \left| TN \right|}{\left| TP \right| + \left| FP \right| + \left| TN \right| + \left| FN \right|}$.

Additionally, the recommendation-based approaches need a threshold value $T$, which is used for the similarity between the predicted name and buggy or fixed names.
It is manually set for the MCC task in prior studies~\cite{cognac_2021,nguyen2020suggesting,deepname}, and following them, we set it as 0.85 in this study.

\section{Experimental Results}
This section reports all the results of our target techniques and answers the research questions.


\subsection{RQ1: Can the target techniques recommend the correct names for a given method body?}
\label{sec.rq1}
The first research question is whether the target techniques listed in Section~\ref{subsec.subject} can perform well for the MNR task when they are evaluated on \benchmark instead of their original datasets.
Following the protocol described in Section~\ref{sec:mnr}, we feed each {\bf MethodBody} in \benchmark to each technique and take the corresponding recommended name as a result.
After getting the results, we compute the performance based on the metrics defined in Section~\ref{sec:mnrmetrics}.

The target techniques show lower performance on \benchmark than in their original studies.
Table~\ref{tab:rq1_results} shows the MNR results as $x/y$, where $x$ is our result and $y$ (in light grey) is the previously reported result.
Note that Spot~\cite{liu2019learning} did not report on the performance of the MNR task with the metrics described in Section~\ref{subsec.metrics}, but rather mainly uses first token matching.
Thus, we cannot directly compare with the original performance of Spot.
Still, Spot performs worst among the target techniques.

\begin{table}[!htp]
\centering
\footnotesize
\caption{Results for the Method Name Recommendation.}
\resizebox{0.7\columnwidth}{!}{%
\begin{tabular}{l|ccc}
\toprule
&
\textbf{\begin{tabular}[c]{@{}c@{}}Spot~\cite{liu2019learning} \end{tabular}}
&
\textbf{\begin{tabular}[c]{@{}c@{}}Cognac~\cite{cognac_2021} \end{tabular}}
& \textbf{\begin{tabular}[c]{@{}c@{}}GTNM~\cite{gtnm_2022} \end{tabular}}
\\ \midrule 
\textbf{EMAcc} &3.50  / \textcolor{gray}{-} & 8.57 / \textcolor{gray}{51.80} & 14.73 / \textcolor{gray}{62.01} \\
\textbf{$\mbox{Precision}_{MNR}$}   &30.79  / \textcolor{gray}{-} & 38.59 / \textcolor{gray}{71.40} & 47.39 /
\textcolor{gray}{77.01} \\
\textbf{$\mbox{Recall}_{MNR}$}      &22.90 / \textcolor{gray}{-} & 25.12 / \textcolor{gray}{61.90} & 41.80 / \textcolor{gray}{74.15} \\
\textbf{$\mbox{F1-score}_{MNR}$}    &26.27 / \textcolor{gray}{-} & 30.49 / \textcolor{gray}{66.30} & 44.42 / \textcolor{gray}{75.60} \\
\bottomrule
\end{tabular}
}
\begin{changemargin}{2.2cm}{2.2cm}
\begin{flushleft}
{The gray values (\textcolor{gray}{nn.nn}) denote the reported results in their papers while `-' indicates that such results are not available.
\vspace{2mm}
}
\end{flushleft}
\end{changemargin}
\label{tab:rq1_results}
\end{table}

Thanks to \benchmark, where all renames are backed by code review (i.e., the {\bf ReviewComment} component), we can figure out the reason why the techniques fail to recommend proper method names. Cognac incorrectly recommends \underline{\it size()} for a method, and the following review\footnote{\url{https://github.com/apache/kafka/pull/1486\#discussion\_r66461876}} corresponds to the case in which {\bf Buggy\-Name} and {\bf FixedName} are \underline{\it size()} and \underline{\it approximateNumEntries()}, respectively:

~\\
\noindent\fcolorbox{gray}{gray!30}{%
\begin{minipage}{0.97\linewidth}{\footnotesize
\underline{\bf Reviewer A}: It seems like this method doesn't return an exact ``size'' for the RocksDb implementation so we need to spec it differently if we want to include it. We should also consider the right name and whether we actually want to expose it.
\\
\ldots
\\
\underline{\bf Reviewer B}: ``approximate'' sounds good. Also to differentiate with number of bytes as @{\it anotherReviewer} pointed out on the ticket, how about use ``approximateNumEntries''?
}%
\end{minipage}}%
~\\

\noindent
According to the review comment corresponding to the method name, the reviewers
suggested ``approximate\ldots'' since the function returns approximated size instead of the exact ``size'' of a specific container.

\benchmark can help improve a method name recommendation technique as it provides the rationale of a method renaming, written by real developers.
For example, the above example suggests that researchers should focus more on semantic and global context information rather than local context. Here is another example:\footnote{\url{https://github.com/apache/kafka/pull/11689\#discussion\_r788921625}} Cognac and GTNM recommends \underline{\it withPartition()} and \underline{\it createTopic()}, respectively, for the following {\bf MethodBody}:

\codeblock{
\begin{lstlisting}[mathescape=true,language=java,linewidth={\linewidth},basicstyle=\scriptsize\ttfamily,numbers=left]
{
        return new TopicPartition(
         record.topic(), record.partition() == null ?
          RecordMetadata.UNKNOWN_PARTITION : record.partition());
}
\end{lstlisting}}

\noindent
while the actual {\bf BuggyName} and {\bf FixedName} in \benchmark are \underline{\it createTopicPartition()} and \underline{\it extract-}
\underline{TopicPartition()}. The corresponding {\bf ReviewComment} is:

~\\
\noindent\fcolorbox{gray}{gray!30}{%
\begin{minipage}{0.97\linewidth}{\footnotesize
\underline{\bf Reviewer}: This (``createTopicPartition'') is not a very good name since ``creating a partition'' usually means calling createPartitions. How about ``extractTopicPartition''?
}%
\end{minipage}}%
~\\

This example motivates that we need to leverage global context such as other source code in the same project since there are similar method implementations
names (e.g., {\it createPartitions()} and {\it createTopics()}) in other source files.

\find {
On \benchmark, the techniques are less successful at recommending appropriate method names (MNR) than in previous evaluations.
Code review comments may provide insights for better name recommendations, such
as exploiting context information available in other methods of the same classes or modules.
}


\subsection{RQ2: Can the target technique detect inconsistent method names?}
\label{sec.rq2}

This research question aims to examine the effectiveness of each target technique with respect to the MCC task described in Section~\ref{sec:mcc}. 
Note that we follow the hypothesis described in a prior study~\cite{liu2019learning} that the methods that have not been commented on in pull requests to be renamed have consistent names.
We feed a pair of a {\bf BuggyName} and the corresponding {\bf MethodBody} in
\benchmark to each technique; we expect that these are classified as
{\it inconsistent} names. In addition, we feed a pair of a {\bf FixedName} and
the corresponding {\bf MethodBody}; we expect that they are
classified as {\it consistent} names.

After constructing a confusion matrix based on the four possible outcomes,
we compute values for the four metrics as described in Section~\ref{sec:mccmetrics}.
The computed values are shown in Table~\ref{tab:rq2_results}.
The results follow the form $x/y$ used in Table~\ref{tab:rq1_results},
where $x$ is the result on \benchmark, and $y$ is the previously reported result.
As there are no previous MCC results for GTNM, the table does not include the previous values in this case.
Nevertheless, it is evident that GTNM can serve as a valuable tool for consistency checking. This is underlined by the fact that a similar methodology has been employed in the development of Cognac, which also aims at name recommendations.


As for the MNR task,
most of the values drop from the original studies except for the case of
$\mbox{Recall}_{IC}$ for Spot. The overall degradation is less than the MNR
task, but some cases are still significant. For example, the $\mbox{Recall}_{C}$
and $\mbox{F1-score}_{C}$ values of
Spot and Cognac drop from 38.20\% and 55.60\% to 5.00\% and 14.25\%,
respectively. 
This indicates that most of the {\bf FixedName} and {\bf MethodBody} pairs are incorrectly classified as {\it inconsistent} ones.
Although the drop of $\mbox{Precision}_{IC}$ and
$\mbox{Precision}_{C}$ is not as significant as the drop in $\mbox{Recall}_{C}$ and $\mbox{F1-score}_{C}$,
the values are just around $\approx50$, which indicates that the performance
is not better than random classification, as the MCC task is fundamentally a binary classification.


\begin{table}[htp!]
\centering
\footnotesize
\caption{Results for Method Name Consistency Checking.}
\resizebox{0.7\columnwidth}{!}{%
\begin{tabular}{cl|ccc}
\toprule
\multicolumn{2}{c|}{}                                 & \textbf{\begin{tabular}[c]{@{}c@{}}Spot~\cite{liu2019learning} \end{tabular}} & \textbf{\begin{tabular}[c]{@{}c@{}}Cognac~\cite{cognac_2021} \end{tabular}} & \textbf{\begin{tabular}[c]{@{}c@{}}GTNM~\cite{gtnm_2022} \end{tabular}} \\ \midrule
\multicolumn{1}{c|}{\textbf{}}   & \textbf{$\mbox{Precision}_{IC}$} & 50.52 / \textcolor{gray}{56.80} & 48.27 / \textcolor{gray}{68.60} & 43.56 / - \\
\multicolumn{1}{c|}{\textbf{IC}} & \textbf{$\mbox{Recall}_{IC}$}    & 97.00 / \textcolor{gray}{84.50} & 80.00 / \textcolor{gray}{97.60} & 66.00 / - \\
\multicolumn{1}{c|}{\textbf{}}   & \textbf{$\mbox{F1-score}_{IC}$}  & 64.44 / \textcolor{gray}{67.90} & 60.21 / \textcolor{gray}{80.60} & 52.49 / - \\
\midrule
\multicolumn{1}{c|}{\textbf{}}   & \textbf{$\mbox{Precision}_{C}$} & 62.50 / \textcolor{gray}{72.00} & 41.61 / \textcolor{gray}{96.00} & 29.90 / - \\
\multicolumn{1}{c|}{\textbf{C}}  & \textbf{$\mbox{Recall}_{C}$}    & 5.00 / \textcolor{gray}{38.20} & 14.25 / \textcolor{gray}{55.60} & 14.50 / - \\
\multicolumn{1}{c|}{\textbf{}}   & \textbf{$\mbox{F1-score}_{C}$}  & 9.26 / \textcolor{gray}{49.90} & 21.23 / \textcolor{gray}{70.40} & 19.53 / - \\
\midrule
\multicolumn{2}{c|}{\textbf{Accuracy}}                & 51.00 / \textcolor{gray}{60.90} & 47.12 / \textcolor{gray}{76.60} & 40.25 / - \\
\bottomrule
\end{tabular}
\footnotesize
}
\begin{changemargin}{2.2cm}{2.2cm}
\begin{flushleft}
{The gray values (\textcolor{gray}{nn.nn}) denote the reported results under their own test oracle, while `-' indicates that such results are not available.}
\vspace{2mm}
\end{flushleft}
\end{changemargin}
\label{tab:rq2_results}
\end{table}


The following example shows the reasons for some failures.
Cognac and GTNM differently classify the following pair of {\bf BuggyName} and {\bf MethodBody} as {\it consistent} and {\it inconsistent}, respectively.

\begin{enumerate}
  \item {\bf BuggyName}: \underline{\it append()}
  \item {\bf MethodBody}: 
  \codeblock{
\begin{lstlisting}[mathescape=true,language=java,linewidth={\linewidth},basicstyle=\scriptsize\ttfamily,numbers=left]
{
    return append(lastOffset < 0 ? baseOffset : lastOffset + 1,
     timestamp, key, value);
}
\end{lstlisting}}
\end{enumerate}
whereas the {\bf FixedName} is \underline{\it appendWithOffset()} corresponding to the {\bf MethodBody}. Both Cognac and GTNM classify the pair of this {\bf FixedName} and the {\bf MethodBody} as {\it inconsistent}. The {\bf ReviewComment}\footnote{\url{https://github.com/apache/kafka/pull/2282\#discussion\_r93328221}} is as follows:

~\\
\noindent\fcolorbox{gray}{gray!30}{%
\begin{minipage}{0.97\linewidth}{\footnotesize
\underline{\bf Reviewer}: Hmm, can we think of another name for the methods that increment the offset automatically?
\\
\ldots
\\
\underline{\bf Committer}: Or maybe this can be the default and the other can be named ``appendWithOffset''?
\\
\ldots
\\
\underline{\bf Reviewer}: That works for me.
}%
\end{minipage}}%
~\\


The techniques fail to correctly classify both the {\bf BuggyName} and the
{\bf FixedName} since they recommend \underline{\it append()} and \underline{\it  addrecord()}
as method names. This indicates that the techniques should better utilize the information available
in {\bf MethodBody} as it mentions ``Offset'' tokens several times. In addition, the reviewers talk about the method incrementing the offset automatically.


\find {
The target techniques achieve a performance similar to random classification as they overclassify consistent names as inconsistent names. The reviews in \benchmark can give intuitions on how to improve the classification performance, such as utilizing more information in the method body.
}


\subsection{RQ3: What happens when we test the effect of balanced vs. imbalanced data?}
\label{sec.rq3}

The objective of this research question is to assess the efficacy of each target technique when evaluating them on
an imbalanced dataset (a huge number of consistent pairs and a small number of inconsistent ones), which reflects the MCC task in real practice,
as outlined in Section~\ref{sec:reality}.
Similar to RQ2, we feed the pairs of method names and bodies.
For inconsistent pairs, we assess all the {\bf BuggyName} and {\bf MethodBody} inconsistent pairs in \benchmark using each technique. 
For consistent pairs, we assess all the {\bf StableName} and  {\bf StableBody} pairs from the dataset described in Section~\ref{sec:reality}; we expect that they will all be classified as {\it consistent}.
This is the `imbalanced dataset' (400 vs. 13,137) in this study.

On the imbalanced dataset, the target techniques mostly fail to classify consistent names correctly.
The results are shown in Table~\ref{tab:rq3_results} and denoted as $x/y$, where $x$ is the result of the imbalanced dataset described above, while $y$ is the result for the balanced dataset (same with Table~\ref{tab:rq2_results}).
$Precision_{IC}$ values are significantly decreased, 
with a decline of 90.40\%, 94.39\%, and 94.90\% for Spot, Cognac, and GTNM, respectively.
As only a few inconsistent pairs are classified as consistent, $Precision_{C}$ values rise, with an increase of 55.82\%, 131.36\%, and 212.41\%, respectively.



\begin{table}[htp!]
\centering
\footnotesize
\caption{Results for Method Name Consistency Checking with the real world reflecting dataset.}
\resizebox{0.7\columnwidth}{!}{%
\begin{tabular}{cl|ccc}
\toprule
\multicolumn{2}{c|}{}                                 & \textbf{\begin{tabular}[c]{@{}c@{}}Spot~\cite{liu2019learning} \end{tabular}} & \textbf{\begin{tabular}[c]{@{}c@{}}Cognac~\cite{cognac_2021}\end{tabular}} & \textbf{\begin{tabular}[c]{@{}c@{}}GTNM~\cite{gtnm_2022} \end{tabular}} \\ \midrule
\multicolumn{1}{c|}{\textbf{}}   & \textbf{$\mbox{Precision}_{IC}$} & 4.85 / \textcolor{gray}{50.52} & 2.71 / \textcolor{gray}{48.27} & 2.22 / \textcolor{gray}{43.56} \\
\multicolumn{1}{c|}{\textbf{IC}} & \textbf{$\mbox{Recall}_{IC}$}    & 97.00 / \textcolor{gray}{97.00} & 80.00 / \textcolor{gray}{80.00} & 66.00 / \textcolor{gray}{66.00} \\
\multicolumn{1}{c|}{\textbf{}}   & \textbf{$\mbox{F1-score}_{IC}$} & 9.24 / \textcolor{gray}{64.44} & 5.25 / \textcolor{gray}{60.21} & 4.30 / \textcolor{gray}{52.49} \\
\midrule
\multicolumn{1}{c|}{\textbf{}}   & \textbf{$\mbox{Precision}_{C}$} & 97.39 / \textcolor{gray}{62.50} & 96.27 / \textcolor{gray}{41.61} & 93.41 / \textcolor{gray}{29.90}\\
\multicolumn{1}{c|}{\textbf{C}}  & \textbf{$\mbox{Recall}_{C}$}    & 5.56 / \textcolor{gray}{5.00} & 15.25 / \textcolor{gray}{14.25} & 14.24 / \textcolor{gray}{14.50}\\
\multicolumn{1}{c|}{\textbf{}}   & \textbf{$\mbox{F1-score}_{C}$}  & 10.52 / \textcolor{gray}{9.26} & 26.32 / \textcolor{gray}{21.23} & 24.72 / \textcolor{gray}{19.53}\\
\midrule
\multicolumn{2}{c|}{\textbf{Accuracy}}                & 9.89 / \textcolor{gray}{51.00} & 17.11 / \textcolor{gray}{47.12} & 15.73 / \textcolor{gray}{40.25}\\
\bottomrule
\end{tabular}
}
\begin{changemargin}{2.2cm}{2.5cm}
\begin{flushleft}
{The gray values (\textcolor{gray}{nn.nn}) denote the previous results from Section~\ref{sec.rq2}.
\vspace{2mm}
}
\end{flushleft}
\end{changemargin}
\label{tab:rq3_results}
\end{table}

The results in Table~\ref{tab:rq3_results} indicate that
the techniques are highly biased to classify method name body pairs as inconsistent ones.
The following pair\footnote{\url{https://github.com/wangpeibin713/flink/blob/748ad82745d9d493922150fe007136e125a50209/flink-table/flink-table-runtime-blink/src/test/java/org/apache/flink/table/runtime/operators/deduplicate/ProcTimeDeduplicateKeepLastRowFunctionTest.java}} shows an example of incorrect classification.
Both Cognac and GTNM classify the following pair of {\bf StableName} and {\bf StableBody} as {\it inconsistent} with different recommendations.
\begin{enumerate}
  \item {\bf StableName}: \underline{\it testWithStateTtlDisabled()}
  \item {\bf StableBody}:
  \codeblock{
\begin{lstlisting}[mathescape=true,language=java,linewidth={\linewidth},basicstyle=\scriptsize\ttfamily,numbers=left]
{
     ProcTimeDeduplicateKeepLastRowFunction func = 
     createFunctionWithoutStateTtl(true, true);
     ...
     testHarness.processElement(insertRecord("book", 1L, 12));
     ...
     // Keep LastRow in deduplicate may send UPDATE_BEFORE
     ...
}
\end{lstlisting}}
\end{enumerate}


\noindent Both Cognac and GTNM fail to correctly recommend {\bf StableName}.
They suggest \underline{\it testKeepLastRow-}
\underline{Function()} and \underline{\it  testInsertWithUpdate()}, respectively. 
This result indicates that the techniques need to further focus on the invoked methods as the key was the method, {\it createFunctionWithoutStateTtl()}. 

Overall, the techniques need to relax the threshold values when detecting inconsistent pairs.
For the following {\bf StableBody},\footnote{\url{https://github.com/open-beagle/shardingsphere/blob/5be6a2c81c647cdffb1f7f17db9c69204be14b4e/shardingsphere-proxy/shardingsphere-proxy-backend/src/test/java/org/apache/shardingsphere/proxy/backend/text/distsql/rdl/resource/AlterResourceBackendHandlerTest.java\#L132}}
Cognac and GTNM recommend {\it getAlterResource()} and {\it createSimpleResourceStatement()}, respectively, which are quite similar to the actual {\bf StableName}, {\it createAlterResourceStatement()}:
\codeblock{
\begin{lstlisting}[mathescape=true,language=java,linewidth={\linewidth},basicstyle=\scriptsize\ttfamily,numbers=left]
{
    return new AlterResourceStatement(Collections.singleton
     (new DataSourceSegment(resourceName, 
     "jdbc:mysql://127.0.0.1:3306/ds_0", 
     null, null, null, "root", "", new Properties())));
}
\end{lstlisting}}

\noindent As the recommended names and {\bf StableName} share many common tokens,
it is appropriate to classify the above case as {\it consistent}, but the techniques
fail to correctly classify it because their threshold values are too tight.

\find{
The  imbalanced dataset, reflecting the real practice, may reveal the shortcomings such as too strong threshold values for inconsistent name classification.
}


\section{Discussion}
\label{sec.discussionfuture}

We mainly discuss the experimental settings for evaluating the target techniques as well as the differences between the datasets. 
The following discussion could also deliver actionable insights for future research directions.

\subsection{Threshold \textit{T} used in the MCC task}
\label{subsec.threshold}
When performing the MCC task, most of the techniques 
compute the similarity between the ground truth name and the predicted name.
This similarity calculation bears a threshold value $T$, which is heuristically decided in the previous studies~\cite{cognac_2021,nguyen2020suggesting,deepname}. 
This threshold value then determines the performance.
However, the choice of $T$ depends on the technique.
For example, MNire~\cite{nguyen2020suggesting} takes various values for $T$ while Cognac~\cite{cognac_2021} uses 0.85 as a fixed value. Moreover, the threshold is chosen to maximize the performance of each technique.

Although the impact of the threshold $T$ is investigated in a literature~\cite{nguyen2020suggesting}, it was only to maximize the performance of each technique.
This indicates that it needs to be tuned depending on the approach and the choice of $T$ potentially causes bias in different scenarios.
Furthermore, one of the prior studies~\cite{cognac_2021} stated, ``we never know a method name is consistent or not before the detection in practice'' regarding the decision on the threshold.
This may imply that their test oracles cannot guarantee the quality for evaluating MCC/MNR tasks.
We investigate the performance of the recommendation-based techniques with $T$ as 1 as we hypothesize {\bf FixedName} in \benchmark may be the ground truth.

\begin{table}[htp!]
\centering
\footnotesize
\caption{The results of Method Name Consistency Checking with the fixed value for threshold $T$ as 1.}
\resizebox{0.6\columnwidth}{!}{%
\begin{tabular}{cl|ccc}
\toprule
\multicolumn{2}{c|}{} &
\textbf{\begin{tabular}[c]{@{}c@{}}Cognac~\cite{cognac_2021} \end{tabular}} & \textbf{\begin{tabular}[c]{@{}c@{}}GTNM~\cite{gtnm_2022} \end{tabular}} \\ \midrule
\multicolumn{1}{c|}{\textbf{}}   & \textbf{$\mbox{Precision}_{IC}$} & 47.98 / \textcolor{gray}{48.27} & 43.49 / \textcolor{gray}{43.56} \\
\multicolumn{1}{c|}{\textbf{IC}} & \textbf{$\mbox{Recall}_{IC}$}    & 80.00 / \textcolor{gray}{80.00} & 66.75 / \textcolor{gray}{66.00} \\
\multicolumn{1}{c|}{\textbf{}}   & \textbf{$\mbox{F1-score}_{IC}$}  & 59.98 / \textcolor{gray}{60.21} & 52.66 / \textcolor{gray}{52.49} \\
\midrule
\multicolumn{1}{c|}{\textbf{}}   & \textbf{$\mbox{Precision}_{C}$} & 39.85 / \textcolor{gray}{41.61} & 28.49 / \textcolor{gray}{29.90}\\
\multicolumn{1}{c|}{\textbf{C}}  & \textbf{$\mbox{Recall}_{C}$}    & 13.25 / \textcolor{gray}{14.25} & 13.25 / \textcolor{gray}{14.50}\\
\multicolumn{1}{c|}{\textbf{}}   & \textbf{$\mbox{F1-score}_{C}$}  & 19.89 / \textcolor{gray}{21.23} & 18.09 / \textcolor{gray}{19.53}\\
\midrule
\multicolumn{2}{c|}{\textbf{Accuracy}}                & 46.62 / \textcolor{gray}{47.12} & 40.00 / \textcolor{gray}{40.25}\\
\bottomrule
\end{tabular}
}
\begin{changemargin}{2.9cm}{2.9cm}
\begin{flushleft}
{The gray values (\textcolor{gray}{nn.nn}) denote the previous results from Section~\ref{sec.rq2} as we aim to discover the effects of the threshold value $T$.
\vspace{2mm}
}
\end{flushleft}
\end{changemargin}
\label{tab:dis_threshold1}
\end{table}


The results in Table~\ref{tab:dis_threshold1} indicate that the target techniques are slightly influenced by the threshold value.
Cognac's recommendations for the $IC$ class do not change at all.
However, there are more false positives for the $C$ class, reducing the precision.
More importantly, both subjects experience a decrease in performance for the $C$ class.
Intriguingly, there is a slight performance increase in GTNM's recommendation on $\mbox{Recall}_{IC}$ class (i.e., 66.00\% to 66.75\%).
The increase in true positives has resulted in this phenomenon, which is unexpected as it is commonly believed that the higher the threshold value, the more difficulty the techniques will encounter.
Inspired by this, we motivate the following discussion.


\subsection{Measurement of True Positives for the Inconsistent Class}
\label{subsec:IC_setting}
During the MCC task of recommendation-based techniques, i.e., Cognac and GTNM, we discovered that the initial evaluation design may be biased due to the following reasons.
A binary classification is not suitable for these approaches, as they initially recommend and then check the inconsistency based on the predictions.
In this way, a recommended method name cannot ensure if a method name is inconsistent or consistent.
Previous studies, such as Cognac~\cite{cognac_2021}, define a true positive in their evaluation as a case where the similarity between the predicted tokens and the oracle tokens from the inconsistent name is less than the threshold value $T$ (i.e., $Sim(o,p)$ < $T$). 
This means that the lower the model's recommendation performance, the more true positives it has for the inconsistent class. 
However, this approach may not accurately reflect the goal of improving method name consistency, as the predicted names should ultimately match the ground truth names.
To address this, we revised the evaluation setting and only consider a predicted name as a true positive if its similarity to the oracle tokens from the consistent name is equal to 1, as this represents a perfect match with the ground truth.

\begin{table}[htp!]
\centering
\footnotesize
\caption{The results of Method Name Consistency Checking with corrected measurement of the true positives for IC class.}
\resizebox{0.6\columnwidth}{!}{%
\begin{tabular}{cl|ccc}
\toprule
\multicolumn{2}{c|}{} &
\textbf{\begin{tabular}[c]{@{}c@{}}Cognac~\cite{cognac_2021}\end{tabular}} & \textbf{\begin{tabular}[c]{@{}c@{}}GTNM~\cite{gtnm_2022}\end{tabular}} \\ \midrule
\multicolumn{1}{c|}{\textbf{}}   & \textbf{$\mbox{Precision}_{IC}$} & 6.79 / \textcolor{gray}{48.27} & 15.56 / \textcolor{gray}{43.56} \\
\multicolumn{1}{c|}{\textbf{IC}} & \textbf{$\mbox{Recall}_{IC}$}    & 6.25 / \textcolor{gray}{80.00} & 15.75 / \textcolor{gray}{66.00} \\
\multicolumn{1}{c|}{\textbf{}}   & \textbf{$\mbox{F1-score}_{IC}$}  & 6.51 / \textcolor{gray}{60.21} & 15.65 / \textcolor{gray}{52.49} \\
\bottomrule
\end{tabular}
}
\begin{changemargin}{2.9cm}{2.9cm}
\begin{flushleft}
{The gray values (\textcolor{gray}{nn.nn}) denote the previous results from Section~\ref{sec.rq2} as we aim to check the phenomenon after the correction.
\vspace{2mm}
}
\end{flushleft}
\end{changemargin}
\label{tab:dis_tp_for_ic}
\end{table}



The findings presented in Table~\ref{tab:dis_tp_for_ic} indicate a marked decrease in performance compared to the initial results.
Cognac decreased by 86\%, 92\%, and 89\% for $Precision_{IC}$, $Recall_{IC}$, and \textit{F1-score}$_{IC}$, respectively.
GTNM performed slightly better, with decreases of 64\%, 76\%, and 70\% on each metric.
The results indicate that recommendation-based approaches demonstrate significant limitations on inconsistency checking.
Overall, there is a need for further improvement in such approaches as the results in Table~\ref{tab:dis_tp_for_ic} do not seem to be promising.
Additionally, such an evaluation protocol requires careful examination.


\subsection{Size of the Test Dataset}
\textbf{On the method name recommendation.}
Most recommendation-based studies~\cite{gtnm_2022,deepname,cognac_2021,nguyen2020suggesting} employ the test dataset from Java-large~\cite{alon2019code2vec} for the MNR task.
This dataset includes approximately 636K test methods from 61K files which is a very large number compared to our datasets.
Although we understand that the bigger the dataset is, the more solid and robust the evaluation that can be conducted, such a dataset has been solely cloned from open-source projects without any filtering criteria.
This suggests that there is no evidence that such method names are ever changed, let alone whether the changes are reviewed by others.


\textbf{On the method name consistency checking.}
Different from the dataset commonly used for the MNR task, most prior techniques~\cite{liu2019learning,li2021namechecker,cognac_2021,nguyen2020suggesting,deepname} leverage the dataset from a checking-based study~\cite{liu2019learning}, which has been constructed using specific filtering criteria.
The filtering criteria include methods whose names have been changed in a commit without any alteration to the body code. 
Additional conditions, such as disregarding method changes with typographical errors and those that do not involve changes to the first sub-token, are also applied as the techniques are sensitive to such cases.
These filtering criteria ensure that the dataset only contains methods whose names have been explicitly changed by developers.
The dataset comprises 2,805 methods from 430 projects for the test phase.
However, it cannot guarantee that the method name changes are associated with the inconsistency of the method name and its body
since there are cases where the developer only changed the method body first and then fixed the name after~\cite{wen2020empirical,wen2022quick}.


\textbf{On our benchmark.}
Our main goal is to conduct an empirical investigation of the existing tools with different perspectives (i.e., code review).
While we continuously collect and craft more method and review pairs, we acknowledge that the current number of test oracles in this study is relatively small compared to the existing ones.
As the ultimate goal of the MCC/MNR tasks is to be as precise and practical as possible, it is vital that the test oracles closely resemble the ground truth. 
To achieve this, we propose utilizing a code review process involving the code reviewers as a means to enhance the quality of the data.
We demonstrate the significant differences in the lengths of names between the method names from our benchmark and those of existing datasets.
This may imply that the benchmark can be leveraged to evaluate MCC/MNR techniques with different scenarios as we described in earlier sections.

In summary, the set of test oracles used for the MNR task is the largest, but its practical usefulness is questionable. 
The dataset used in the MCC task cannot guarantee whether the method name changes are associated with the inconsistency checking task or coincident code changes as it is collected without considering the rationale behind them.
Meanwhile, our benchmark dataset is composed of methods that have been reviewed and explicitly pointed out by code reviewers of open-source projects.
As we continually extend the benchmark, we expect it will allow for a more comprehensive evaluation of MCC/MNR techniques.



\subsection{Threats to Validity}

\textbf{Threats to external validity}
may lie on the targeted three techniques~\cite{liu2019learning,cognac_2021,gtnm_2022} that are open-source; thus, the results may not be representative of other techniques or closed-source techniques.
Furthermore, as we consider approaches targeting Java, our conclusions are only valid for this language.
A threat may lie in the dataset as well. We utilize a common dataset for training all the recommendation-based techniques, and such a dataset can possibly be biased, which could lead the models in the wrong direction.
To mitigate this threat, we incorporate a large-scale dataset~\cite{alon2018code2seq} that consists of high-quality and well-maintained open-source projects.
The identifier names in these large-scale projects are known to be mostly consistent~\cite{alon2018code2seq,alon2019code2vec}.
Additionally, such approaches focus on learning common features from the majority of the methods. 
As a result, during the training phase, a natural mitigation of this concern could be attained.

\noindent
\textbf{Threats to internal validity}
may include human-reviewed method names in our benchmark. 
To address this threat, the authors cross-checked the method names and bodies.
We found that there exist very long method names with huge bodies which could be ambiguous to decide whether the names are consistent or not. 
All the authors who cross-checked the method names agreed on the final version of the benchmark.
In addition, although it is not guaranteed that all positive items have been identified, appropriate rationales for the name changes could support the quality of the naming dataset as well as recommendation tasks.

\noindent
\textbf{Threats to construct validity}
relate to the evaluation metrics we employed.
We took the same metrics as target techniques~\cite{liu2019learning,cognac_2021,gtnm_2022}, which are Accuracy, Precision, Recall, F1-score, over sub-tokens, and EMAcc.
Although there may exist other potential metrics to measure the performance of MCC/MNR techniques, evaluating such techniques with the prior metrics over sub-tokens is the approach taken by all the related literature.


\section{Related Work}
\label{sec.relatedwork}

\textbf{Empirical Investigation on Identifier Naming.}
Based on the assumption, \textit{poor names make programs harder to understand and maintain}, many researchers~\cite{takang1996effects,liblit2006cognitive,butler2009relating,butler2011mining} have conducted empirical investigations on naming source code identifiers.
The majority of these studies explore the impact of names on source code readability~\cite{scalabrino2018comprehensive,fakhoury2018effect}, program comprehension~\cite{hofmeister2017shorter,scalabrino2019automatically,wainakh2021idbench}, and software maintainability~\cite{butler2010exploring}.
In addition, several studies~\cite{higo2012often,kim2016automatic} investigate the unintended inconsistency of the identifiers.
Recently, researchers~\cite{jiang2019machine,alsuhaibani_2021_2021} have begun to focus on naming methods, as these are the smallest unit of an aggregated functionality.
Such studies investigate existing method name recommendation techniques or survey professional developers to understand the impact of naming choices.

\noindent
\textbf{MCC/MNR techniques, on the applied models.}
Approaches were built based on textual similarity and co-occurrence.
H{\o}st and {\O}stvold~\cite{host_debugging_2009} introduced a debugging approach for method names by inferring naming rules based on the return type, control flow, and method parameters.
Lucia et al.~\cite{de2010improving} proposed an information retrieval based approach
using LSI~\cite{deerwester1990indexing} to compute the textual similarity
to improve program comprehension.

Recently, researchers have proposed learning-based techniques.
Allamanis et al.~\cite{allamanis2015suggesting,allamanis2016convolutional} employ deep learning techniques (i.e., basic and copy convolutional attention models) to recommend method names.
Liu et al.~\cite{liu2019learning} take the same embedding strategy with Paragraph Vector~\cite{le2014distributed} and convolutional neural networks (CNN)~\cite{lecun1999object} considering additional code nodes at the abstract syntax tree (AST) level to capture code semantic information~\cite{mou2016convolutional}.
Code2seq~\cite{alon2018code2seq} and Code2vec~\cite{alon2019code2vec} are well-known code embedding techniques that are evaluated with the method name recommendation task.
Code2seq~\cite{alon2018code2seq} follows the standard encoder-decoder architecture~\cite{cho2014learning} with a bi-directional LSTM~\cite{huang2015bidirectional} while Code2vec~\cite{alon2019code2vec} leverages a newly designed path-attention network.
Wang et al.~\cite{cognac_2021} further added the pointer-generator network~\cite{see2017get} and introduced Cognac aiming to avoid the out-of-vocabulary problem.
With a similar concept, Nguyen et al.~\cite{nguyen2020suggesting} and Li et al.~\cite{deepname} proposed MNire and DeepName, respectively, considering different contexts. These models leverage the RNN-based Seq2seq~\cite{britz2017massive} model with attention mechanism~\cite{bahdanau2014neural,luong2014addressing} aiming to capture contextual sentences.
Liu et al.~\cite{gtnm_2022}'s GTNM is a transformer-based model that leverages the self-attention mechanism capturing rich semantic dependencies.

\noindent
\textbf{MCC/MNR techniques, on the leveraged contexts.}
To check and recommend the method names, various related contexts are considered.
Most approaches exploit the features from the local context, such as the method implementation (e.g., tokens of the return type, parameters, control flow graph, data flow graph, AST, and each identifier in the body)~\cite{suzuki2014approach,allamanis2015suggesting,allamanis2016convolutional,alon2018code2seq,alon2019code2vec,liu2019learning}.
Recently, researchers started to explore a wider range for the context, such as the enclosing class name~\cite{nguyen2020suggesting} and method invocation relations (i.e., caller and callee)~\cite{cognac_2021,deepname}.
Liu et al.~\cite{gtnm_2022} further extended the context to project-specific information (i.e., in-file methods and cross-file contextual methods) and documentation of the method (i.e., Javadoc)~\cite{gtnm_2022}.
A few studies leverage high-level artifacts such as software requirement documents~\cite{de2010improving} and semantic profiles of the method implementations~\cite{host_debugging_2009}.


\section{Conclusion}
\label{sec.conclusion}
We investigated existing method name consistency checking and recommendation approaches with a novel benchmark that contains clear rationales for changing method names.
Investigating the recent empirical studies on developer activities for naming identifiers, we discovered that developers could finalize the method names with code review comments.
Inspired by this phenomenon, we construct a benchmark for method name changes by collecting and mapping with the related review comments, which can be a further layer of checking the consistency.
We hold the perspective that our benchmark could serve as a reference for method name consistency checking and recommendation techniques. This is because the enclosed test oracles are derived from recommendations or insights provided by code reviewers within the projects.
To validate existing studies on the benchmark, we explored whether they show consistent performance compared to the previous dataset that does not rely on clear rationales.
Although there was an exceptional case due to the scalability issue with an existing technique, Spot, we carefully selected our target techniques and design the experimental setting for a fair comparison.
The results of the existing approaches demonstrate a consistent decrease in our test oracles for most of the metrics they defined.
Overall, this indicates that there still exists room for improvement, especially when they encounter different scenarios and evaluation protocols.
Moreover, we revealed that there exist potentially biased features in the evaluation of existing techniques which research should consider to fill the gap between research and practice.
We hope that these actionable insights could be applied to future research directions.




\section{Data Availability}
We publicly release a replication package that includes all the code and datasets to reproduce the experiments of our study at \url{https://figshare.com/s/8cdb4e3208e01991e45c}

\bibliographystyle{ACM-Reference-Format}


\begin{thebibliography}{54}


\ifx \showCODEN    \undefined \def \showCODEN     #1{\unskip}     \fi
\ifx \showDOI      \undefined \def \showDOI       #1{#1}\fi
\ifx \showISBNx    \undefined \def \showISBNx     #1{\unskip}     \fi
\ifx \showISBNxiii \undefined \def \showISBNxiii  #1{\unskip}     \fi
\ifx \showISSN     \undefined \def \showISSN      #1{\unskip}     \fi
\ifx \showLCCN     \undefined \def \showLCCN      #1{\unskip}     \fi
\ifx \shownote     \undefined \def \shownote      #1{#1}          \fi
\ifx \showarticletitle \undefined \def \showarticletitle #1{#1}   \fi
\ifx \showURL      \undefined \def \showURL       {\relax}        \fi
\providecommand\bibfield[2]{#2}
\providecommand\bibinfo[2]{#2}
\providecommand\natexlab[1]{#1}
\providecommand\showeprint[2][]{arXiv:#2}

\bibitem[Aha(2013)]%
        {aha2013lazy}
\bibfield{author}{\bibinfo{person}{David~W Aha}.}
  \bibinfo{year}{2013}\natexlab{}.
\newblock \bibinfo{booktitle}{\emph{Lazy learning}}.
\newblock \bibinfo{publisher}{Springer Science \& Business Media}.
\newblock


\bibitem[Allamanis et~al\mbox{.}(2015)]%
        {allamanis2015suggesting}
\bibfield{author}{\bibinfo{person}{Miltiadis Allamanis},
  \bibinfo{person}{Earl~T Barr}, \bibinfo{person}{Christian Bird}, {and}
  \bibinfo{person}{Charles Sutton}.} \bibinfo{year}{2015}\natexlab{}.
\newblock \showarticletitle{Suggesting accurate method and class names}. In
  \bibinfo{booktitle}{\emph{Proceedings of the 2015 10th joint meeting on
  foundations of software engineering}}. \bibinfo{pages}{38--49}.
\newblock


\bibitem[Allamanis et~al\mbox{.}(2016)]%
        {allamanis2016convolutional}
\bibfield{author}{\bibinfo{person}{Miltiadis Allamanis}, \bibinfo{person}{Hao
  Peng}, {and} \bibinfo{person}{Charles Sutton}.}
  \bibinfo{year}{2016}\natexlab{}.
\newblock \showarticletitle{A convolutional attention network for extreme
  summarization of source code}. In \bibinfo{booktitle}{\emph{International
  conference on machine learning}}. PMLR, \bibinfo{pages}{2091--2100}.
\newblock


\bibitem[Alon et~al\mbox{.}(2019a)]%
        {alon2018code2seq}
\bibfield{author}{\bibinfo{person}{Uri Alon}, \bibinfo{person}{Shaked Brody},
  \bibinfo{person}{Omer Levy}, {and} \bibinfo{person}{Eran Yahav}.}
  \bibinfo{year}{2019}\natexlab{a}.
\newblock \showarticletitle{code2seq: Generating Sequences from Structured
  Representations of Code}. In \bibinfo{booktitle}{\emph{International
  Conference on Learning Representations}}.
\newblock
\urldef\tempurl%
\url{https://openreview.net/forum?id=H1gKYo09tX}
\showURL{%
\tempurl}


\bibitem[Alon et~al\mbox{.}(2019b)]%
        {alon2019code2vec}
\bibfield{author}{\bibinfo{person}{Uri Alon}, \bibinfo{person}{Meital
  Zilberstein}, \bibinfo{person}{Omer Levy}, {and} \bibinfo{person}{Eran
  Yahav}.} \bibinfo{year}{2019}\natexlab{b}.
\newblock \showarticletitle{code2vec: Learning distributed representations of
  code}.
\newblock \bibinfo{journal}{\emph{Proceedings of the ACM on Programming
  Languages}} \bibinfo{volume}{3}, \bibinfo{number}{POPL}
  (\bibinfo{year}{2019}), \bibinfo{pages}{1--29}.
\newblock


\bibitem[Alsuhaibani et~al\mbox{.}(2021)]%
        {alsuhaibani_2021_2021}
\bibfield{author}{\bibinfo{person}{Reem Alsuhaibani},
  \bibinfo{person}{Christian Newman}, \bibinfo{person}{Michael Decker},
  \bibinfo{person}{Michael Collard}, {and} \bibinfo{person}{Jonathan Maletic}.}
  \bibinfo{year}{2021}\natexlab{}.
\newblock \showarticletitle{2021 [{ICSE}]\_On the {Naming} of {Methods}: {A}
  {Survey} of {Professional} {Developers}}. In \bibinfo{booktitle}{\emph{2021
  {IEEE}/{ACM} 43rd {International} {Conference} on {Software} {Engineering}
  ({ICSE})}}. \bibinfo{publisher}{IEEE}, \bibinfo{address}{Madrid, ES},
  \bibinfo{pages}{587--599}.
\newblock
\showISBNx{978-1-66540-296-5}
\urldef\tempurl%
\url{https://doi.org/10.1109/ICSE43902.2021.00061}
\showDOI{\tempurl}


\bibitem[Arnaoudova et~al\mbox{.}(2016)]%
        {arnaoudova2016linguistic}
\bibfield{author}{\bibinfo{person}{Venera Arnaoudova},
  \bibinfo{person}{Massimiliano Di~Penta}, {and} \bibinfo{person}{Giuliano
  Antoniol}.} \bibinfo{year}{2016}\natexlab{}.
\newblock \showarticletitle{Linguistic antipatterns: What they are and how
  developers perceive them}.
\newblock \bibinfo{journal}{\emph{Empirical Software Engineering}}
  \bibinfo{volume}{21}, \bibinfo{number}{1} (\bibinfo{year}{2016}),
  \bibinfo{pages}{104--158}.
\newblock


\bibitem[Arnaoudova et~al\mbox{.}(2014)]%
        {arnaoudova2014repent}
\bibfield{author}{\bibinfo{person}{Venera Arnaoudova}, \bibinfo{person}{Laleh~M
  Eshkevari}, \bibinfo{person}{Massimiliano Di~Penta}, \bibinfo{person}{Rocco
  Oliveto}, \bibinfo{person}{Giuliano Antoniol}, {and}
  \bibinfo{person}{Yann-Ga{\"e}l Gu{\'e}h{\'e}neuc}.}
  \bibinfo{year}{2014}\natexlab{}.
\newblock \showarticletitle{Repent: Analyzing the nature of identifier
  renamings}.
\newblock \bibinfo{journal}{\emph{IEEE Transactions on Software Engineering}}
  \bibinfo{volume}{40}, \bibinfo{number}{5} (\bibinfo{year}{2014}),
  \bibinfo{pages}{502--532}.
\newblock


\bibitem[Bahdanau et~al\mbox{.}(2014)]%
        {bahdanau2014neural}
\bibfield{author}{\bibinfo{person}{Dzmitry Bahdanau},
  \bibinfo{person}{Kyunghyun Cho}, {and} \bibinfo{person}{Yoshua Bengio}.}
  \bibinfo{year}{2014}\natexlab{}.
\newblock \showarticletitle{Neural machine translation by jointly learning to
  align and translate}.
\newblock \bibinfo{journal}{\emph{arXiv preprint arXiv:1409.0473}}
  (\bibinfo{year}{2014}).
\newblock


\bibitem[Britz et~al\mbox{.}(2017)]%
        {britz2017massive}
\bibfield{author}{\bibinfo{person}{Denny Britz}, \bibinfo{person}{Anna Goldie},
  \bibinfo{person}{Minh-Thang Luong}, {and} \bibinfo{person}{Quoc Le}.}
  \bibinfo{year}{2017}\natexlab{}.
\newblock \showarticletitle{Massive exploration of neural machine translation
  architectures}.
\newblock \bibinfo{journal}{\emph{arXiv preprint arXiv:1703.03906}}
  (\bibinfo{year}{2017}).
\newblock


\bibitem[Butler et~al\mbox{.}(2009)]%
        {butler2009relating}
\bibfield{author}{\bibinfo{person}{Simon Butler}, \bibinfo{person}{Michel
  Wermelinger}, \bibinfo{person}{Yijun Yu}, {and} \bibinfo{person}{Helen
  Sharp}.} \bibinfo{year}{2009}\natexlab{}.
\newblock \showarticletitle{Relating identifier naming flaws and code quality:
  An empirical study}. In \bibinfo{booktitle}{\emph{2009 16th Working
  Conference on Reverse Engineering}}. IEEE, \bibinfo{pages}{31--35}.
\newblock


\bibitem[Butler et~al\mbox{.}(2010)]%
        {butler2010exploring}
\bibfield{author}{\bibinfo{person}{Simon Butler}, \bibinfo{person}{Michel
  Wermelinger}, \bibinfo{person}{Yijun Yu}, {and} \bibinfo{person}{Helen
  Sharp}.} \bibinfo{year}{2010}\natexlab{}.
\newblock \showarticletitle{Exploring the influence of identifier names on code
  quality: An empirical study}. In \bibinfo{booktitle}{\emph{2010 14th European
  Conference on Software Maintenance and Reengineering}}. IEEE,
  \bibinfo{pages}{156--165}.
\newblock


\bibitem[Butler et~al\mbox{.}(2011a)]%
        {butler2011improving}
\bibfield{author}{\bibinfo{person}{Simon Butler}, \bibinfo{person}{Michel
  Wermelinger}, \bibinfo{person}{Yijun Yu}, {and} \bibinfo{person}{Helen
  Sharp}.} \bibinfo{year}{2011}\natexlab{a}.
\newblock \showarticletitle{Improving the tokenisation of identifier names}. In
  \bibinfo{booktitle}{\emph{European Conference on Object-Oriented
  Programming}}. Springer, \bibinfo{pages}{130--154}.
\newblock


\bibitem[Butler et~al\mbox{.}(2011b)]%
        {butler2011mining}
\bibfield{author}{\bibinfo{person}{Simon Butler}, \bibinfo{person}{Michel
  Wermelinger}, \bibinfo{person}{Yijun Yu}, {and} \bibinfo{person}{Helen
  Sharp}.} \bibinfo{year}{2011}\natexlab{b}.
\newblock \showarticletitle{Mining java class naming conventions}. In
  \bibinfo{booktitle}{\emph{2011 27th IEEE International Conference on Software
  Maintenance (ICSM)}}. IEEE, \bibinfo{pages}{93--102}.
\newblock


\bibitem[Cho et~al\mbox{.}(2014)]%
        {cho2014learning}
\bibfield{author}{\bibinfo{person}{Kyunghyun Cho}, \bibinfo{person}{Bart
  Van~Merri{\"e}nboer}, \bibinfo{person}{Caglar Gulcehre},
  \bibinfo{person}{Dzmitry Bahdanau}, \bibinfo{person}{Fethi Bougares},
  \bibinfo{person}{Holger Schwenk}, {and} \bibinfo{person}{Yoshua Bengio}.}
  \bibinfo{year}{2014}\natexlab{}.
\newblock \showarticletitle{Learning phrase representations using RNN
  encoder-decoder for statistical machine translation}.
\newblock \bibinfo{journal}{\emph{arXiv preprint arXiv:1406.1078}}
  (\bibinfo{year}{2014}).
\newblock


\bibitem[De~Lucia et~al\mbox{.}(2010)]%
        {de2010improving}
\bibfield{author}{\bibinfo{person}{Andrea De~Lucia},
  \bibinfo{person}{Massimiliano Di~Penta}, {and} \bibinfo{person}{Rocco
  Oliveto}.} \bibinfo{year}{2010}\natexlab{}.
\newblock \showarticletitle{Improving source code lexicon via traceability and
  information retrieval}.
\newblock \bibinfo{journal}{\emph{IEEE Transactions on Software Engineering}}
  \bibinfo{volume}{37}, \bibinfo{number}{2} (\bibinfo{year}{2010}),
  \bibinfo{pages}{205--227}.
\newblock


\bibitem[Deerwester et~al\mbox{.}(1990)]%
        {deerwester1990indexing}
\bibfield{author}{\bibinfo{person}{Scott Deerwester}, \bibinfo{person}{Susan~T
  Dumais}, \bibinfo{person}{George~W Furnas}, \bibinfo{person}{Thomas~K
  Landauer}, {and} \bibinfo{person}{Richard Harshman}.}
  \bibinfo{year}{1990}\natexlab{}.
\newblock \showarticletitle{Indexing by latent semantic analysis}.
\newblock \bibinfo{journal}{\emph{Journal of the American society for
  information science}} \bibinfo{volume}{41}, \bibinfo{number}{6}
  (\bibinfo{year}{1990}), \bibinfo{pages}{391--407}.
\newblock


\bibitem[Fakhoury et~al\mbox{.}(2018)]%
        {fakhoury2018effect}
\bibfield{author}{\bibinfo{person}{Sarah Fakhoury}, \bibinfo{person}{Yuzhan
  Ma}, \bibinfo{person}{Venera Arnaoudova}, {and} \bibinfo{person}{Olusola
  Adesope}.} \bibinfo{year}{2018}\natexlab{}.
\newblock \showarticletitle{The effect of poor source code lexicon and
  readability on developers' cognitive load}. In \bibinfo{booktitle}{\emph{2018
  IEEE/ACM 26th International Conference on Program Comprehension (ICPC)}}.
  IEEE, \bibinfo{pages}{286--28610}.
\newblock


\bibitem[GitHub(2022)]%
        {github}
GitHub \bibinfo{year}{2022}\natexlab{}.
\newblock \bibinfo{title}{https://github.com/}.
\newblock
\newblock


\bibitem[Gu et~al\mbox{.}(2018)]%
        {gu2018deep}
\bibfield{author}{\bibinfo{person}{Xiaodong Gu}, \bibinfo{person}{Hongyu
  Zhang}, {and} \bibinfo{person}{Sunghun Kim}.}
  \bibinfo{year}{2018}\natexlab{}.
\newblock \showarticletitle{Deep code search}. In
  \bibinfo{booktitle}{\emph{2018 IEEE/ACM 40th International Conference on
  Software Engineering (ICSE)}}. IEEE, \bibinfo{pages}{933--944}.
\newblock


\bibitem[Hastie et~al\mbox{.}(2009)]%
        {hastie2009unsupervised}
\bibfield{author}{\bibinfo{person}{Trevor Hastie}, \bibinfo{person}{Robert
  Tibshirani}, {and} \bibinfo{person}{Jerome Friedman}.}
  \bibinfo{year}{2009}\natexlab{}.
\newblock \showarticletitle{Unsupervised learning}.
\newblock In \bibinfo{booktitle}{\emph{The elements of statistical learning}}.
  \bibinfo{publisher}{Springer}, \bibinfo{pages}{485--585}.
\newblock


\bibitem[Higo and Kusumoto(2012)]%
        {higo2012often}
\bibfield{author}{\bibinfo{person}{Yoshiki Higo} {and} \bibinfo{person}{Shinji
  Kusumoto}.} \bibinfo{year}{2012}\natexlab{}.
\newblock \showarticletitle{How often do unintended inconsistencies happen?
  Deriving modification patterns and detecting overlooked code fragments}. In
  \bibinfo{booktitle}{\emph{2012 28th IEEE International Conference on Software
  Maintenance (ICSM)}}. IEEE, \bibinfo{pages}{222--231}.
\newblock


\bibitem[Hofmeister et~al\mbox{.}(2017)]%
        {hofmeister2017shorter}
\bibfield{author}{\bibinfo{person}{Johannes Hofmeister}, \bibinfo{person}{Janet
  Siegmund}, {and} \bibinfo{person}{Daniel~V Holt}.}
  \bibinfo{year}{2017}\natexlab{}.
\newblock \showarticletitle{Shorter identifier names take longer to
  comprehend}. In \bibinfo{booktitle}{\emph{2017 IEEE 24th International
  conference on software analysis, evolution and reengineering (SANER)}}. IEEE,
  \bibinfo{pages}{217--227}.
\newblock


\bibitem[H{\o}st and {\O}stvold(2009)]%
        {host2009debugging}
\bibfield{author}{\bibinfo{person}{Einar~W H{\o}st} {and}
  \bibinfo{person}{Bjarte~M {\O}stvold}.} \bibinfo{year}{2009}\natexlab{}.
\newblock \showarticletitle{Debugging method names}. In
  \bibinfo{booktitle}{\emph{European Conference on Object-Oriented
  Programming}}. Springer, \bibinfo{pages}{294--317}.
\newblock


\bibitem[Huang et~al\mbox{.}(2015)]%
        {huang2015bidirectional}
\bibfield{author}{\bibinfo{person}{Zhiheng Huang}, \bibinfo{person}{Wei Xu},
  {and} \bibinfo{person}{Kai Yu}.} \bibinfo{year}{2015}\natexlab{}.
\newblock \showarticletitle{Bidirectional LSTM-CRF models for sequence
  tagging}.
\newblock \bibinfo{journal}{\emph{arXiv preprint arXiv:1508.01991}}
  (\bibinfo{year}{2015}).
\newblock


\bibitem[Høst and Østvold(2009)]%
        {host_debugging_2009}
\bibfield{author}{\bibinfo{person}{Einar~W. Høst} {and}
  \bibinfo{person}{Bjarte~M. Østvold}.} \bibinfo{year}{2009}\natexlab{}.
\newblock \showarticletitle{Debugging {Method} {Names}}. In
  \bibinfo{booktitle}{\emph{{ECOOP} 2009 – {Object}-{Oriented}
  {Programming}}}, \bibfield{editor}{\bibinfo{person}{Sophia Drossopoulou}}
  (Ed.). \bibinfo{publisher}{Springer Berlin Heidelberg},
  \bibinfo{address}{Berlin, Heidelberg}, \bibinfo{pages}{294--317}.
\newblock
\showISBNx{978-3-642-03013-0}


\bibitem[Jiang et~al\mbox{.}(2019)]%
        {jiang2019machine}
\bibfield{author}{\bibinfo{person}{Lin Jiang}, \bibinfo{person}{Hui Liu}, {and}
  \bibinfo{person}{He Jiang}.} \bibinfo{year}{2019}\natexlab{}.
\newblock \showarticletitle{Machine learning based recommendation of method
  names: how far are we}. In \bibinfo{booktitle}{\emph{2019 34th IEEE/ACM
  International Conference on Automated Software Engineering (ASE)}}. IEEE,
  \bibinfo{pages}{602--614}.
\newblock


\bibitem[Karlton(2022)]%
        {karlton}
Karlton \bibinfo{year}{2022}\natexlab{}.
\newblock \bibinfo{title}{https://www.nndb.com/people/400/000031307/}.
\newblock
\newblock


\bibitem[Kim and Kim(2016)]%
        {kim2016automatic}
\bibfield{author}{\bibinfo{person}{Suntae Kim} {and} \bibinfo{person}{Dongsun
  Kim}.} \bibinfo{year}{2016}\natexlab{}.
\newblock \showarticletitle{Automatic identifier inconsistency detection using
  code dictionary}.
\newblock \bibinfo{journal}{\emph{Empirical Software Engineering}}
  \bibinfo{volume}{21}, \bibinfo{number}{2} (\bibinfo{year}{2016}),
  \bibinfo{pages}{565--604}.
\newblock


\bibitem[Lawrie et~al\mbox{.}(2006)]%
        {lawrie2006s}
\bibfield{author}{\bibinfo{person}{Dawn Lawrie}, \bibinfo{person}{Christopher
  Morrell}, \bibinfo{person}{Henry Feild}, {and} \bibinfo{person}{David
  Binkley}.} \bibinfo{year}{2006}\natexlab{}.
\newblock \showarticletitle{What's in a Name? A Study of Identifiers}. In
  \bibinfo{booktitle}{\emph{14th IEEE international conference on program
  comprehension (ICPC'06)}}. IEEE, \bibinfo{pages}{3--12}.
\newblock


\bibitem[Le and Mikolov(2014)]%
        {le2014distributed}
\bibfield{author}{\bibinfo{person}{Quoc Le} {and} \bibinfo{person}{Tomas
  Mikolov}.} \bibinfo{year}{2014}\natexlab{}.
\newblock \showarticletitle{Distributed representations of sentences and
  documents}. In \bibinfo{booktitle}{\emph{International conference on machine
  learning}}. PMLR, \bibinfo{pages}{1188--1196}.
\newblock


\bibitem[LeCun et~al\mbox{.}(1999)]%
        {lecun1999object}
\bibfield{author}{\bibinfo{person}{Yann LeCun}, \bibinfo{person}{Patrick
  Haffner}, \bibinfo{person}{L{\'e}on Bottou}, {and} \bibinfo{person}{Yoshua
  Bengio}.} \bibinfo{year}{1999}\natexlab{}.
\newblock \showarticletitle{Object recognition with gradient-based learning}.
\newblock In \bibinfo{booktitle}{\emph{Shape, contour and grouping in computer
  vision}}. \bibinfo{publisher}{Springer}, \bibinfo{pages}{319--345}.
\newblock


\bibitem[Li et~al\mbox{.}(2021a)]%
        {9712079}
\bibfield{author}{\bibinfo{person}{Kejun Li}, \bibinfo{person}{Taiming Wang},
  {and} \bibinfo{person}{Hui Liu}.} \bibinfo{year}{2021}\natexlab{a}.
\newblock \showarticletitle{NameChecker: Detecting Inconsistency between Method
  Names and Method Bodies}. In \bibinfo{booktitle}{\emph{2021 28th Asia-Pacific
  Software Engineering Conference (APSEC)}}. \bibinfo{pages}{22--31}.
\newblock
\urldef\tempurl%
\url{https://doi.org/10.1109/APSEC53868.2021.00010}
\showDOI{\tempurl}


\bibitem[Li et~al\mbox{.}(2021b)]%
        {li2021namechecker}
\bibfield{author}{\bibinfo{person}{Kejun Li}, \bibinfo{person}{Taiming Wang},
  {and} \bibinfo{person}{Hui Liu}.} \bibinfo{year}{2021}\natexlab{b}.
\newblock \showarticletitle{NameChecker: Detecting Inconsistency between Method
  Names and Method Bodies}. In \bibinfo{booktitle}{\emph{2021 28th Asia-Pacific
  Software Engineering Conference (APSEC)}}. IEEE, \bibinfo{pages}{22--31}.
\newblock


\bibitem[Li et~al\mbox{.}(2021c)]%
        {deepname}
\bibfield{author}{\bibinfo{person}{Yi Li}, \bibinfo{person}{Shaohua Wang},
  {and} \bibinfo{person}{Tien Nguyen}.} \bibinfo{year}{2021}\natexlab{c}.
\newblock \showarticletitle{A Context-based Automated Approach for Method Name
  Consistency Checking and Suggestion}. In \bibinfo{booktitle}{\emph{2021
  IEEE/ACM 43rd International Conference on Software Engineering (ICSE)}}.
  IEEE, \bibinfo{pages}{574--586}.
\newblock


\bibitem[Liblit et~al\mbox{.}(2006)]%
        {liblit2006cognitive}
\bibfield{author}{\bibinfo{person}{Ben Liblit}, \bibinfo{person}{Andrew Begel},
  {and} \bibinfo{person}{Eve Sweetser}.} \bibinfo{year}{2006}\natexlab{}.
\newblock \showarticletitle{Cognitive Perspectives on the Role of Naming in
  Computer Programs.}. In \bibinfo{booktitle}{\emph{PPIG}}. Citeseer,
  \bibinfo{pages}{11}.
\newblock


\bibitem[Liu et~al\mbox{.}(2022)]%
        {gtnm_2022}
\bibfield{author}{\bibinfo{person}{F. Liu}, \bibinfo{person}{G. Li},
  \bibinfo{person}{Z. Fu}, \bibinfo{person}{S. Lu}, \bibinfo{person}{Y. Hao},
  {and} \bibinfo{person}{Z. Jin}.} \bibinfo{year}{2022}\natexlab{}.
\newblock \showarticletitle{Learning to Recommend Method Names with Global
  Context}. In \bibinfo{booktitle}{\emph{2022 IEEE/ACM 44th International
  Conference on Software Engineering (ICSE)}}. \bibinfo{publisher}{IEEE
  Computer Society}, \bibinfo{address}{Los Alamitos, CA, USA},
  \bibinfo{pages}{1294--1306}.
\newblock
\urldef\tempurl%
\url{https://doi.org/10.1145/3510003.3510154}
\showDOI{\tempurl}


\bibitem[Liu et~al\mbox{.}(2019)]%
        {liu2019learning}
\bibfield{author}{\bibinfo{person}{Kui Liu}, \bibinfo{person}{Dongsun Kim},
  \bibinfo{person}{Tegawend{\'e}~F Bissyand{\'e}}, \bibinfo{person}{Taeyoung
  Kim}, \bibinfo{person}{Kisub Kim}, \bibinfo{person}{Anil Koyuncu},
  \bibinfo{person}{Suntae Kim}, {and} \bibinfo{person}{Yves Le~Traon}.}
  \bibinfo{year}{2019}\natexlab{}.
\newblock \showarticletitle{Learning to spot and refactor inconsistent method
  names}. In \bibinfo{booktitle}{\emph{2019 IEEE/ACM 41st International
  Conference on Software Engineering (ICSE)}}. IEEE, \bibinfo{pages}{1--12}.
\newblock


\bibitem[Luong et~al\mbox{.}(2014)]%
        {luong2014addressing}
\bibfield{author}{\bibinfo{person}{Minh-Thang Luong}, \bibinfo{person}{Ilya
  Sutskever}, \bibinfo{person}{Quoc~V Le}, \bibinfo{person}{Oriol Vinyals},
  {and} \bibinfo{person}{Wojciech Zaremba}.} \bibinfo{year}{2014}\natexlab{}.
\newblock \showarticletitle{Addressing the rare word problem in neural machine
  translation}.
\newblock \bibinfo{journal}{\emph{arXiv preprint arXiv:1410.8206}}
  (\bibinfo{year}{2014}).
\newblock


\bibitem[Maletic and Marcus(2001)]%
        {maletic2001supporting}
\bibfield{author}{\bibinfo{person}{Jonathan~I Maletic} {and}
  \bibinfo{person}{Andrian Marcus}.} \bibinfo{year}{2001}\natexlab{}.
\newblock \showarticletitle{Supporting program comprehension using semantic and
  structural information}. In \bibinfo{booktitle}{\emph{Proceedings of the 23rd
  International Conference on Software Engineering. ICSE 2001}}. IEEE,
  \bibinfo{pages}{103--112}.
\newblock


\bibitem[Mou et~al\mbox{.}(2016)]%
        {mou2016convolutional}
\bibfield{author}{\bibinfo{person}{Lili Mou}, \bibinfo{person}{Ge Li},
  \bibinfo{person}{Lu Zhang}, \bibinfo{person}{Tao Wang}, {and}
  \bibinfo{person}{Zhi Jin}.} \bibinfo{year}{2016}\natexlab{}.
\newblock \showarticletitle{Convolutional neural networks over tree structures
  for programming language processing}. In \bibinfo{booktitle}{\emph{Thirtieth
  AAAI conference on artificial intelligence}}.
\newblock


\bibitem[Nguyen et~al\mbox{.}(2020)]%
        {nguyen2020suggesting}
\bibfield{author}{\bibinfo{person}{Son Nguyen}, \bibinfo{person}{Hung Phan},
  \bibinfo{person}{Trinh Le}, {and} \bibinfo{person}{Tien~N Nguyen}.}
  \bibinfo{year}{2020}\natexlab{}.
\newblock \showarticletitle{Suggesting natural method names to check name
  consistencies}. In \bibinfo{booktitle}{\emph{Proceedings of the ACM/IEEE 42nd
  international conference on software engineering}}.
  \bibinfo{pages}{1372--1384}.
\newblock


\bibitem[Ribeiro et~al\mbox{.}(2016)]%
        {ribeiro2016should}
\bibfield{author}{\bibinfo{person}{Marco~Tulio Ribeiro},
  \bibinfo{person}{Sameer Singh}, {and} \bibinfo{person}{Carlos Guestrin}.}
  \bibinfo{year}{2016}\natexlab{}.
\newblock \showarticletitle{" Why should i trust you?" Explaining the
  predictions of any classifier}. In \bibinfo{booktitle}{\emph{Proceedings of
  the 22nd ACM SIGKDD international conference on knowledge discovery and data
  mining}}. \bibinfo{pages}{1135--1144}.
\newblock


\bibitem[Scalabrino et~al\mbox{.}(2019)]%
        {scalabrino2019automatically}
\bibfield{author}{\bibinfo{person}{Simone Scalabrino},
  \bibinfo{person}{Gabriele Bavota}, \bibinfo{person}{Christopher Vendome},
  \bibinfo{person}{Mario Linares-Vasquez}, \bibinfo{person}{Denys Poshyvanyk},
  {and} \bibinfo{person}{Rocco Oliveto}.} \bibinfo{year}{2019}\natexlab{}.
\newblock \showarticletitle{Automatically assessing code understandability}.
\newblock \bibinfo{journal}{\emph{IEEE Transactions on Software Engineering}}
  \bibinfo{volume}{47}, \bibinfo{number}{3} (\bibinfo{year}{2019}),
  \bibinfo{pages}{595--613}.
\newblock


\bibitem[Scalabrino et~al\mbox{.}(2018)]%
        {scalabrino2018comprehensive}
\bibfield{author}{\bibinfo{person}{Simone Scalabrino}, \bibinfo{person}{Mario
  Linares-V{\'a}squez}, \bibinfo{person}{Rocco Oliveto}, {and}
  \bibinfo{person}{Denys Poshyvanyk}.} \bibinfo{year}{2018}\natexlab{}.
\newblock \showarticletitle{A comprehensive model for code readability}.
\newblock \bibinfo{journal}{\emph{Journal of Software: Evolution and Process}}
  \bibinfo{volume}{30}, \bibinfo{number}{6} (\bibinfo{year}{2018}),
  \bibinfo{pages}{e1958}.
\newblock


\bibitem[Schankin et~al\mbox{.}(2018)]%
        {schankin2018descriptive}
\bibfield{author}{\bibinfo{person}{Andrea Schankin}, \bibinfo{person}{Annika
  Berger}, \bibinfo{person}{Daniel~V Holt}, \bibinfo{person}{Johannes~C
  Hofmeister}, \bibinfo{person}{Till Riedel}, {and} \bibinfo{person}{Michael
  Beigl}.} \bibinfo{year}{2018}\natexlab{}.
\newblock \showarticletitle{Descriptive compound identifier names improve
  source code comprehension}. In \bibinfo{booktitle}{\emph{2018 IEEE/ACM 26th
  International Conference on Program Comprehension (ICPC)}}. IEEE,
  \bibinfo{pages}{31--3109}.
\newblock


\bibitem[See et~al\mbox{.}(2017)]%
        {see2017get}
\bibfield{author}{\bibinfo{person}{Abigail See}, \bibinfo{person}{Peter~J Liu},
  {and} \bibinfo{person}{Christopher~D Manning}.}
  \bibinfo{year}{2017}\natexlab{}.
\newblock \showarticletitle{Get to the point: Summarization with
  pointer-generator networks}.
\newblock \bibinfo{journal}{\emph{arXiv preprint arXiv:1704.04368}}
  (\bibinfo{year}{2017}).
\newblock


\bibitem[Suzuki et~al\mbox{.}(2014)]%
        {suzuki2014approach}
\bibfield{author}{\bibinfo{person}{Takayuki Suzuki}, \bibinfo{person}{Kazunori
  Sakamoto}, \bibinfo{person}{Fuyuki Ishikawa}, {and} \bibinfo{person}{Shinichi
  Honiden}.} \bibinfo{year}{2014}\natexlab{}.
\newblock \showarticletitle{An approach for evaluating and suggesting method
  names using n-gram models}. In \bibinfo{booktitle}{\emph{Proceedings of the
  22nd International Conference on Program Comprehension}}.
  \bibinfo{pages}{271--274}.
\newblock


\bibitem[Takang et~al\mbox{.}(1996)]%
        {takang1996effects}
\bibfield{author}{\bibinfo{person}{Armstrong~A Takang},
  \bibinfo{person}{Penny~A Grubb}, {and} \bibinfo{person}{Robert~D Macredie}.}
  \bibinfo{year}{1996}\natexlab{}.
\newblock \showarticletitle{The effects of comments and identifier names on
  program comprehensibility: an experimental investigation}.
\newblock \bibinfo{journal}{\emph{J. Prog. Lang.}} \bibinfo{volume}{4},
  \bibinfo{number}{3} (\bibinfo{year}{1996}), \bibinfo{pages}{143--167}.
\newblock


\bibitem[Vaswani et~al\mbox{.}(2017)]%
        {vaswani2017attention}
\bibfield{author}{\bibinfo{person}{Ashish Vaswani}, \bibinfo{person}{Noam
  Shazeer}, \bibinfo{person}{Niki Parmar}, \bibinfo{person}{Jakob Uszkoreit},
  \bibinfo{person}{Llion Jones}, \bibinfo{person}{Aidan~N Gomez},
  \bibinfo{person}{{\L}ukasz Kaiser}, {and} \bibinfo{person}{Illia
  Polosukhin}.} \bibinfo{year}{2017}\natexlab{}.
\newblock \showarticletitle{Attention is all you need}.
\newblock \bibinfo{journal}{\emph{Advances in neural information processing
  systems}}  \bibinfo{volume}{30} (\bibinfo{year}{2017}).
\newblock


\bibitem[Wainakh et~al\mbox{.}(2021)]%
        {wainakh2021idbench}
\bibfield{author}{\bibinfo{person}{Yaza Wainakh}, \bibinfo{person}{Moiz Rauf},
  {and} \bibinfo{person}{Michael Pradel}.} \bibinfo{year}{2021}\natexlab{}.
\newblock \showarticletitle{Idbench: Evaluating semantic representations of
  identifier names in source code}. In \bibinfo{booktitle}{\emph{2021 IEEE/ACM
  43rd International Conference on Software Engineering (ICSE)}}. IEEE,
  \bibinfo{pages}{562--573}.
\newblock


\bibitem[Wang et~al\mbox{.}(2021)]%
        {cognac_2021}
\bibfield{author}{\bibinfo{person}{Shangwen Wang}, \bibinfo{person}{Ming Wen},
  \bibinfo{person}{Bo Lin}, {and} \bibinfo{person}{Xiaoguang Mao}.}
  \bibinfo{year}{2021}\natexlab{}.
\newblock \showarticletitle{Lightweight global and local contexts guided method
  name recommendation with prior knowledge}. In
  \bibinfo{booktitle}{\emph{Proceedings of the 29th ACM Joint Meeting on
  European Software Engineering Conference and Symposium on the Foundations of
  Software Engineering}}. \bibinfo{pages}{741--753}.
\newblock


\bibitem[Wen et~al\mbox{.}(2020)]%
        {wen2020empirical}
\bibfield{author}{\bibinfo{person}{Fengcai Wen}, \bibinfo{person}{Csaba Nagy},
  \bibinfo{person}{Michele Lanza}, {and} \bibinfo{person}{Gabriele Bavota}.}
  \bibinfo{year}{2020}\natexlab{}.
\newblock \showarticletitle{An empirical study of quick remedy commits}. In
  \bibinfo{booktitle}{\emph{Proceedings of the 28th International Conference on
  Program Comprehension}}. \bibinfo{pages}{60--71}.
\newblock


\bibitem[Wen et~al\mbox{.}(2022)]%
        {wen2022quick}
\bibfield{author}{\bibinfo{person}{Fengcai Wen}, \bibinfo{person}{Csaba Nagy},
  \bibinfo{person}{Michele Lanza}, {and} \bibinfo{person}{Gabriele Bavota}.}
  \bibinfo{year}{2022}\natexlab{}.
\newblock \showarticletitle{Quick remedy commits and their impact on mining
  software repositories}.
\newblock \bibinfo{journal}{\emph{Empirical Software Engineering}}
  \bibinfo{volume}{27} (\bibinfo{year}{2022}), \bibinfo{pages}{1--31}.
\newblock


\end{thebibliography}


\end{document}